%% file: main.tex
\documentclass[acmsmall]{acmart}

\input{preamble/packages}
\input{preamble/acm_preamble}
\begin{document}

\title{A Survey of Graph Pre-processing Methods: From Algorithmic to Hardware Perspectives}

\input{tex/author}

\begin{abstract}

\input{tex/abstract}
\end{abstract}

\input{tex/CCS_concepts}


\maketitle

\section{Introduction}
\label{introduction}
\input{tex/introduction}

\section{Preliminary}
\label{preliminary}
\input{tex/preliminary}

\section{Graph Pre-processing: Tackling Challenges in Graph Processing}
\label{challenge}
\input{tex/challenge}

\section{Graph Pre-processing: Taxonomy with Double-Level Decision}
\label{taxonomy}
\input{tex/taxonomy}

\section{Graph Pre-processing in Algorithmic Perspective}
\label{category-algorithm}
\input{tex/category-algorithm}

\section{Graph Pre-processing in Hardware Perspective}
\label{category-hardware}
\input{tex/category-hardware}

\section{Summary and Comparison}
\label{comparsion}
\input{tex/comparison}

\section{Challenges and Future Directions}
\label{future}
\input{tex/future}

\section{Conclusion}
\label{conclusion}
\input{tex/conclusion}

\begin{acks}
\input{tex/acknowledgement}
\end{acks}

\bibliographystyle{ref/ACM-Reference-Format}

\bibliography{ref/application-survey, ref/method}

\end{document}

%% file: preamble/packages.tex
\usepackage{graphicx}
\usepackage{hyperref}
\usepackage{float}
\usepackage{amsmath}
\usepackage{booktabs}
\usepackage[ruled,vlined,linesnumbered]{algorithm2e}
\usepackage{mathrsfs}
 
\usepackage{amssymb,amsfonts}
\usepackage{color}
\usepackage{multirow}
\usepackage{makecell}
\usepackage{marvosym}
\usepackage{pifont}
\usepackage{soul}
\usepackage{tikz}
\usetikzlibrary{calc}
\usepackage{multibib}
\usepackage{threeparttable}
\usepackage{pifont}
\usepackage[section]{placeins}
\usepackage{arydshln}
\usepackage{caption}

\newcommand \blackCircle[1]{%
  \begin{tikzpicture}[baseline=-0.5ex, inner sep=0pt]
    \draw[fill=black] (0,0) circle (1ex);
    \node[anchor=center, inner sep=0pt, white, font=\small, text height=1.5ex, text depth=0.5ex] at (0,0) {\strut #1};
  \end{tikzpicture}%
}
\newcommand \blackCirclef{%
  \begin{tikzpicture}[baseline=-0.5ex, inner sep=0pt]
    \draw[fill=black] (0,0) circle (1ex);
    \node[anchor=center, inner sep=0pt, white, font=\small, text height=1.5ex, text depth=0.25ex] at (0,0) {f};
  \end{tikzpicture}%
}
\newcommand \blackCircleg{%
  \begin{tikzpicture}[baseline=-0.5ex, inner sep=0pt]
    \draw[fill=black] (0,0) circle (1ex);
    \node[anchor=center, inner sep=0pt, white, font=\small, text height=1.5ex, text depth=0.8ex] at (0,0) {g};
  \end{tikzpicture}%
}


%% file: preamble/acm_preamble.tex
\AtBeginDocument{%
  \providecommand\BibTeX{{%
    \normalfont B\kern-0.5em{\scshape i\kern-0.25em b}\kern-0.8em\TeX}}}

%% file: tex/author.tex

\author{Zhengyang Lv}
\email{lvzhengyang19@mails.ucas.ac.cn}
\orcid{0009-0002-3718-4520}
\author{Mingyu Yan}
\authornotemark[1]
\email{yanmingyu@ict.ac.cn}
\orcid{0000-0002-6915-955X}
\author{Xin Liu}
\email{liuxin196@mails.ucas.ac.cn}
\affiliation{%
  \institution{SKLP, ICT, CAS}
  \streetaddress{No.6 Kexueyuan South Road Zhongguancun,Haidian District}
  \city{Beijing}
  \country{China}
  \postcode{100190}
}
\affiliation{%
  \institution{UCAS}
  \city{Beijing}
  \country{China}
}
\author{Mengyao Dong}
\email{dongmy@shanghaitech.edu.cn}
\affiliation{%
  \institution{ShanghaiTech Univ.}
  \city{Shanghai}
  \country{China}
  }
\affiliation{%
\institution{SHIC}
\city{Shanghai}
\country{China}
}
  
\author{Xiaochun Ye}
\email{yexiaochun@ict.ac.cn}
\author{Dongrui Fan}
\email{Fandr@ict.ac.cn}
\author{Ninghui Sun}
\email{snh@ict.ac.cn}
\affiliation{%
  \institution{ICT, CAS}
  \city{Beijing}
  \country{China}
}


%% file: tex/abstract.tex
Graph-related applications have experienced significant growth in academia and industry, driven by the powerful representation capabilities of graph. However, efficiently executing these applications faces various challenges, such as load imbalance, random memory access, etc. To address these challenges, researchers have proposed various acceleration systems, including software frameworks and hardware accelerators, all of which incorporate graph pre-processing (GPP). GPP serves as a preparatory step before the formal execution of applications, involving techniques such as sampling, reorder, etc. However, GPP execution often remains overlooked, as the primary focus is directed towards enhancing graph applications themselves. This oversight is concerning, especially considering the explosive growth of real-world graph data, where GPP becomes essential and even dominates system running overhead. Furthermore, GPP methods exhibit significant variations across devices and applications due to high customization. Unfortunately, no comprehensive work systematically summarizes GPP. To address this gap and foster a better understanding of GPP, we present a comprehensive survey dedicated to this area. We propose a double-level taxonomy of GPP, considering both algorithmic and hardware perspectives. Through listing relavent works, we illustrate our taxonomy and conduct a thorough analysis and summary of diverse GPP techniques. Lastly, we discuss challenges in GPP and potential future directions.

%% file: tex/CCS_concepts.tex
\ccsdesc[500]{General and reference~Surveys and overviews}
\ccsdesc[500]{Theory of computation~Graph algorithms analysis}
\ccsdesc[500]{Hardware~Hardware accelerators}
\ccsdesc[500]{Computer systems organization~Architectures}

%% file: tex/introduction.tex
\begin{figure}[!t]
    \centering
    \setlength{\abovecaptionskip}{0.2cm}     
    \setlength{\belowcaptionskip}{-0.3cm}   
    \includegraphics[width=\textwidth]{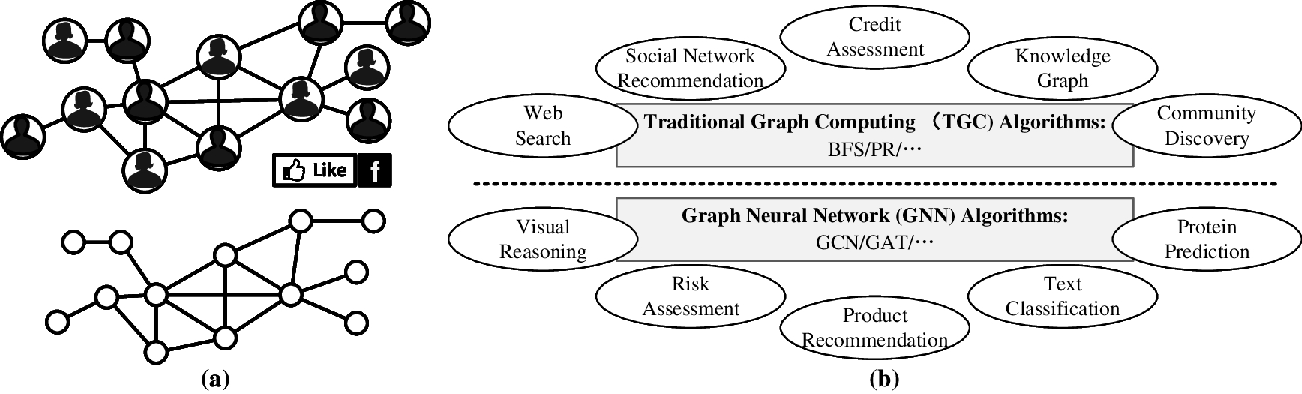}
    \caption{Graph processing algorithms as enablers of many applications that depend on graph data: (a) Graphs abstracted from Facebook networks, where vertices denote users and edges denote users' relationships; (b) Applications of graph processing algorithms.}
    \label{fig:facebook_application}
    \vspace{-0.3cm}  
\end{figure}

Graph processing applications have garnered significant attention for their ability to provide valuable insights from graph data. In various real-world scenarios, data can be effectively represented using graph structures, with social networks being a prime example \cite{Pan2020comprehensive}. For instance, Figure \ref{fig:facebook_application} (a) depicts the abstraction of the social network of Facebook as a graph. There are two most widely used types of graph processing applications: traditional graph computing (TGC), which includes algorithms like breadth-first search (BFS), pagerank (PR), and others; and graph neural network (GNN) such as graph convolution network (GCN) and graph attention network (GAT). These graph processing algorithms find extensive use in various scenarios, including social network recommendation \cite{2019rankingPagerank}, knowledge graph analysis \cite{2019bfsknowledgeGraph}, protein prediction \cite{2021protein-predict}, visual reasoning \cite{2022visual-question-answering}, and more. To handle the exponentially growing scale of graph data, these algorithms have become increasingly popular and are widely deployed in diverse data centers, such as Google-Maps \cite{2021GoogleMaps}, Microsoft-Academic-Graph \cite{2016MicrosoftAcademicGraph}, Alibaba’s E-commerce platform \cite{2019Aligraph}, Baidu-Maps \cite{2020constgat-baidumap}, etc.

The execution of graph processing algorithms faces several challenges, and numerous efforts have been designed to alleviate these issues. \textit{Firstly}, in TGC algorithms, the execution behavior, including factors like resource utilization and operation sequencing, often exhibits irregularities. These irregularities arise from the irregular topology of graphs, leading to irregular workloads, memory access, and communication \cite{che2013pannotia}. To tackle these challenges, various frameworks based on general hardware platforms (CPU \& GPU) have been proposed, such as GraphChi \cite{2012graphchi} and CuSha \cite{2014cusha}. Also, custom architectures have been developed for further acceleration, such as Graphicionado \cite{2016graphicionado} based on ASIC (Application-Specific Integrated Circuit), ForeGraph \cite{2017foregraph} based on FPGA (Field Programmable Gate Array), and GraphR \cite{2018GraphR} based on PIM (Processing-In-Memory). \textit{Secondly}, GNN algorithms exhibit a combination of irregular and regular execution behaviors \cite{2020Characterizing-GCN-GPU,2022characterizing-HGNN-GPU}, as the adding of neural networks (NNs) to transform high-dimensional feature vectors for each vertex. To address both regular and irregular characteristics in GNNs, several dedicated acceleration platforms have been proposed, such as HyGCN \cite{2020HyGCN} based on ASIC and GraphACT \cite{2020GraphACT} based on FPGA. \textit{Thirdly}, the explosive growth of graph data has prompted exploration into parallel processing for large-scale graphs. TGC parallel frameworks include Pregel \cite{2010Pregel}, GraphLab \cite{2012graphlab}, and GNN parallel frameworks include DistDGL \cite{2020DistDGL}, PaGraph \cite{2020PaGraph}, etc.

Graph processing execution heavily relies on a critical operation—graph pre-processing (GPP). For example, GraphChi \cite{2012graphchi}, Graphicionado \cite{2016graphicionado}, GraphDynS \cite{2019GraphDynS}, FPGP \cite{2016fpgp}, and HyGCN \cite{2020HyGCN} utilize reorganization techniques to pre-split the graph data into shards, enabling contiguous memory access and enhancing performance. In parallel graph processing systems like Pregel \cite{2010Pregel}, GraphLab \cite{2012graphlab}, DistDGL \cite{2020DistDGL}, and PaGraph \cite{2020PaGraph}, graph partition is performed in advance to divide large-scale graph data into multiple subgraphs and assign them to multiple processors/machines, achieving load balance and minimizing communication overhead. To facilitate the efficient training of GNN in parallel, PaGraph \cite{2020PaGraph} and DistDGL \cite{2020DistDGL} create mini-batches using sampling techniques. GraphACT \cite{2020GraphACT} and GCNInfer \cite{2020GcnInference} pre-merge common neighbors to reduce subsequent redundant operations. Therefore, GPP is critical for efficiently executing graph processing algorithms, benefiting a wide range of graph processing systems, including single-machine graph processing frameworks, distributed graph processing frameworks, graph processing accelerators, etc.

\begin{figure}[!t]
    \centering
    \setlength{\abovecaptionskip}{0.2cm}     
    \setlength{\belowcaptionskip}{-0.3cm}   
    \includegraphics[width=1\textwidth]{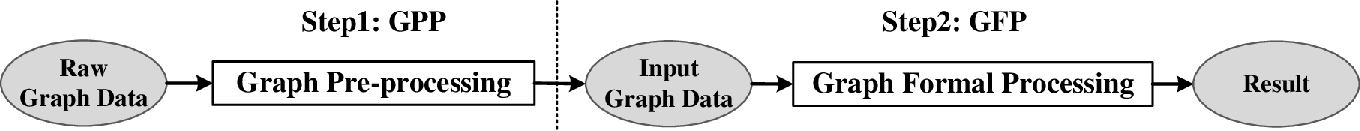}
    \caption{Two main steps in graph processing systems: graph pre-processing (GPP) and graph formal processing (GFP).}
    \label{fig:pipline}
    \vspace{-0.3cm}  
\end{figure}

To provide clarity, we abstract a typical graph processing system into two main steps: graph pre-processing (GPP) and graph formal processing (GFP), as illustrated in Figure \ref{fig:pipline}. During the GPP step, various operations are performed on the raw graph data to prepare the input dataset for subsequent execution of graph processing algorithms. In the GFP step, the computing unit loads the pre-processed data and executes the graph processing algorithm to obtain the final result. It is worth noting that the choice of GPP methods depends on the characteristics of the raw graph dataset, as well as the execution platforms. For example, partition is employed to manage large-scale graph data in parallel systems, with reasearches like DistDGL \cite{2020DistDGL} using CPU-clusters and PaGraph \cite{2020PaGraph} using multi-GPUs. GraphACT \cite{2020GraphACT} employs reconstruction methods to reduce redundant computing on FPGA, achieving high performance and energy efficiency. Overall, GPP offers two main benefits: a) reducing overhead in computing, storage, and communication; b) meeting the execution requirements of various algorithms on devices with limited resources.

Unfortunately, the GPP overhead has become increasingly significant due to the explosive growth of graph data. Next, we give the following examples to visualize the importance of GPP via numerical comparisons. In Graph500 competition~\footnote{https://graph500.org/?page\_id=834}, Fugaku \cite{2020Fugaku}, a petascale supercomputer, exhibits the high GPP time (C\_TIME) of 390 seconds, a striking 1560$\times$ contrast to the BFS execution time of 0.25 seconds. In Gorder \cite{2016Gorder}, graph reorder on a large Twitter dataset takes 1.5 hours, while PageRank completes 100 iterations in just 13.65 minutes. Therefore, if the input graph is not frequently reused, the significant GPP time for large datasets may not be a worthwhile investment. Similarly, in Graphite \cite{2022Graphite}, when executing GraphSAGE, the sampling time comprises over 80\% of the total training time. These examples highlight the significance of reducing GPP overhead to improve the overall execution efficiency of graph processing systems.

Based on the preceding analysis, two significant conclusions emerge, highlighting the pressing need of GPP surveys. Firstly, GPP is crucial for efficient graph processing. Secondly, the GPP overhead is becoming significant and non-negligible, necessitating the reduction of GPP overhead. As a result, the field of GPP holds immense potential for further exploration, and more in-depth studies are required to fully explore the optimization possibilities offered by GPP.

Nevertheless, there is a lack of comprehensive review of GPP techniques across the field, despite some studies analyzing individual GPP methods \cite{2018partitionstream, 2020surveyPartition, 2018NNCompressionSurvey, 2021liu-sampling-survey}. This gap hinders a holistic understanding of the potential optimizations that can be achieved through GPP. In Table \ref{table:survey}, we present a list of related surveys in the field of graph processing and involving GPP methods in these works. Some surveys have explored both TGC acceleration and GNN acceleration techniques, with some touching on GPP methods. For example, surveys on GPU and FPGA-based TGC \cite{2018GraphGPUSurvey, 2019GraphFPGASurvey} involve partition techniques for handling large graphs. Other works \cite{2015ThinkingVertexSurvey, 2018SurveyScalable, 2019GraphAccSurvey, 2021graphMemorySurvey} analyze static and dynamic graph partition in distributed systems and memory-based graph processing systems. Recent surveys \cite{2022GCNFPGAsurvey,2021GNNAlgToAccSurvey,2022bottleneckDGNN,2021GNNTrainingSurvey-full-mini,2023DistGNN-DNN-Survey,2022DistributedGNNSurvey,linhaiyang2022DistributedGNNSurvey,2022surveyGNNAcc-liu,2023surveyGNN-Algo-System-hardware} extensively cover GNN acceleration and describe the significance of GPP steps in GNN execution. However, these surveys still concentrate on analyzing the optimization of GFP step, with GPP not being their primary focus.

\begin{table*}[h]
\vspace{-0.3cm}  
\caption{Surveys about graph processing acceleration and the GPP methods involved}
\vspace{-0.3cm}  
\label{table:survey}
\centering
\small
\begin{tabular}{|m{8.85cm}|c|c|}
\hline
\makecell[c]{\textbf{Survey (Year)}} & \textbf{Domain} & \textbf{GPP}\\ \hline
$\bullet$ Graph Processing on GPUs: A Survey (2018) \cite{2018GraphGPUSurvey} &  \makecell{TGC\\GPU} & \makecell{Partition\\Reorganization} \\ \hline

$\bullet$ Graph Processing on FPGAs: Taxonomy, Survey, Challenges (2019) \cite{2019GraphFPGASurvey} & \makecell{TGC\\FPGA} & \makecell{Partition\\Reorganization} \\ \hline

$\bullet$ A Survey on Graph Processing Accelerators: Challenges and Opportunities (2019) \cite{2019GraphAccSurvey} & \makecell{TGC\\Hardware} & \makecell{Partition\\Reorder}\\ \hline

$\bullet$ Graph Processing and Machine Learning Architectures with Emerging Memory Technologies: A Survey (2021) \cite{2021graphMemorySurvey} & \makecell{TGC \& GNN \\ PIM} & \makecell{Reorganization}\\ \hline

$\bullet$ Thinking Like a Vertex: A Survey of Vertex-Centric Frameworks for Large-Scale Distributed Graph Processing (2015) \cite{2015ThinkingVertexSurvey} & \multirow{3}{*}{\begin{tabular}[c]{@{}c@{}}TGC \\ Parallel System\end{tabular}} & \multirow{3}{*}{\begin{tabular}[c]{@{}c@{}}Partition\\Reorganization\end{tabular}}\\
$\bullet$ Scalable Graph Processing Frameworks: A Taxonomy and Open Challenges (2018) \cite{2018SurveyScalable} &  &  \\ \hline

$\bullet$ A Survey of Field Programmable Gate Array (FPGA)-based Graph Convolutional Neural Network Accelerators: Challenges and Opportunities (2022) \cite{2022GCNFPGAsurvey} & \makecell{GNN\\FPGA} & \makecell{Partition\\Sampling\\Quantification} \\ \hline

$\bullet$ Bottleneck Analysis of Dynamic Graph Neural Network Inference on CPU and GPU (2022) \cite{2022bottleneckDGNN} & \makecell{Dynamic GNN\\ CPU \& GPU} & \makecell{Partition\\Sampling}\\ \hline

$\bullet$ Scalable Graph Neural Network Training: The Case for Sampling (2021) \cite{2021GNNTrainingSurvey-full-mini} & \makecell{GNN\\Algorithm} & \makecell{Partition\\Sampling} \\ \hline


$\bullet$ A Comprehensive Survey on Distributed Training of Graph Neural Networks (2022) \cite{linhaiyang2022DistributedGNNSurvey} & \multirow{2}{*}{\begin{tabular}[c]{@{}c@{}} GNN \\ CPU \& GPU\end{tabular}} & \multirow{2}{*}{\begin{tabular}[c]{@{}c@{}} Partition\\Sampling\end{tabular}} \\
$\bullet$ Distributed Graph Neural Network Training: A Survey (2022) \cite{2022DistributedGNNSurvey} &  &  \\ \hline

$\bullet$ The Evolution of Distributed Systems for Graph Neural Networks and their Origin in Graph Processing and Deep Learning: A Survey (2023) \cite{2023DistGNN-DNN-Survey} & \makecell{GNN\\Parallel System} & \makecell{Partition\\Sampling} \\\hline

$\bullet$ Survey on Graph Neural Network Acceleration: An Algorithmic Perspective (2022) \cite{2022surveyGNNAcc-liu} & \makecell{GNN \\ Algorithm} & \makecell{Many}\\ \hline

$\bullet$ Computing Graph Neural Networks: A Survey from Algorithms to Accelerators (2021) \cite{2021GNNAlgToAccSurvey} & \multirow{3}{*}{\begin{tabular}[c]{@{}c@{}}GNN\\Algorithm\\ Hardware\end{tabular}} & \multirow{3}{*}{Many}\\
$\bullet$ A Survey on Graph Neural Network Acceleration: Algorithms, Systems, and Customized Hardware (2023) \cite{2023surveyGNN-Algo-System-hardware} &  &  \\ \hline

$\bullet$ Survey and Taxonomy of Lossless Graph Compression and Space-Efficient Graph Representations (2018) \cite{2018NNCompressionSurvey} & \makecell{Graph\\Algorithm} & \makecell{Quantification} \\ \hline
$\bullet$ Streaming Graph Partitioning: An Experimental Study (2018) \cite{2018partitionstream} & \multirow{3}{*}{\begin{tabular}[c]{@{}c@{}}GNN\\Algorithm\end{tabular}} & \multirow{3}{*}{Partition}\\
$\bullet$ A Survey of Current Challenges in Partitioning and Processing of Graph Structured Data in Parallel and Distributed Systems (2020) \cite{2020surveyPartition} &  &  \\ \hline

$\bullet$ Sampling Methods For Efficient Training of Graph Convolutional Networks: A Survey (2021) \cite{2021liu-sampling-survey} & \makecell{GNN \\ Algorithm} & \makecell{Sampling}\\ \hline
\end{tabular}
\vspace{-0.5cm}  
\end{table*}

To harness the full potential of GPP in graph processing, it is crucial to perform both hardware and algorithm optimizations. However, there is a gap between hardware acceleration and algorithm optimization in GPP. Existing research predominantly focuses on hardware acceleration of GFP, with limited attention given to GPP, or it may only analyze individual GPP techniques at the algorithmic level. This article primarily aims to bridge this gap by providing a systematic and comprehensive summary and analysis of GPP methods, encompassing both algorithmic and hardware perspectives. We are honored to present a comprehensive overview of GPP methods, with the aim of contributing to the advancement of GPP and offering a reference for further research in this area. Our work may provide valuable insights for the future optimization of GPP execution and graph processing acceleration. Our contributions are as follows:


\textbf{Review}: We review the challenges associated with graph processing execution, considering the aspects of computing, storage, and communication. We highlight the significance of GPP for optimizing execution through relevant examples.

\textbf{Taxonomy}: We classify existing GPP methods and propose a double-level taxonomy from both algorithmic and hardware perspective. The algorithmic categories include graph representation optimization and data representation optimization. The hardware categories encompass efficient computing, storage, and communication.

\textbf{Analysis}: We provide detailed introductions to the existing GPP methods in accordance with the proposed taxonomy. Specifically, we list and analyze related works from both algorithmic and hardware perspectives.

\textbf{Comparison}: We offer a comprehensive summary and comparison of existing GPP methods considering both algorithmic and hardware aspects, allowing for a better understanding of their strengths and weaknesses.

\textbf{Discussion}: We discuss the challenges associated with GPP such as high overhead, accuracy loss, etc. Finally, we outline potential research directions for future exploration.

The rest of this paper is organized as follows: Section \ref{preliminary} provides preliminary GPP information, covering graph concepts and algorithms. Section \ref{challenge} explores execution challenges of graph processing and demonstrates how GPP can address them. Section \ref{taxonomy} presents our double-level GPP taxonomy based on algorithmic optimization factors and hardware optimization effects. Sections \ref{category-algorithm} and \ref{category-hardware} respectively analyze GPP methods with examples from algorithmic and hardware perspectives. Section \ref{comparsion} offers a comprehensive summary and comparison. Section \ref{future} discusses prevailing GPP bottlenecks and potential research directions. Finally, Section \ref{conclusion} concludes our work.

%% file: tex/preliminary.tex
In this section, we will begin by introducing the fundamental concepts of graphs that are used throughout the subsequent sections, covering graph representation and storage format. Then, we will outline the two types of graph processing algorithms: TGC algorithms and GNN algorithms. To facilitate a comprehensive analysis of the execution process for these algorithms, we will also present the programming models that are commonly employed. 

\begin{table*}[h]
\vspace{-0.2cm}  
\caption{Notations and corresponding descriptions used in this work.}
\label{table:notation}
\centering
\small
\vspace{-0.3cm}  
\begin{tabular}{|c|c|}
\hline
Notations                             & Descriptions                                        \\ 
\hline
\textit{G} = (\textit{V}, \textit{E})                     & A graph and its vertex sets \textit{V} and edge sets \textit{E}.                                                 \\
\textit{n}, \textit{m}                                    & The number of vertices and edges, \textit{n} = $\mid$\textit{V}$\mid$, \textit{m} = $\mid$\textit{E}$\mid$.      \\
\textit{v}, \textit{$e_{ij}$}                             & A vertex \textit{v} $\in$ \textit{V}, an edge from \textit{i} to \textit{j}, \textit{$e_{ij}$} $\in$ \textit{E}.  \\
\textit{N}(\textit{v}), \textit{SN}(\textit{v})           & Original and sampled neighbourhood set of vertex \textit{v}.                \\

\textbf{A}, $\widetilde{\textbf{A}}$                      & The original and normalized adjacency matrix of graph.                       \\
\textbf{D}, $\widetilde{\textbf{D}}$                      & The degree matrix of \textbf{A} and $\widetilde{\textbf{A}}$.                \\
$\textbf{W}_{\textit{V}}$, $\textbf{W}_{\textit{E}}$      & The vertex and edge weight matrix of graph.                                  \\
\textit{d}                                                & The dimension of a vertex feature vector or hidden vertex feature vector.    \\
$\textbf{X}, \textbf{H}$                                  & The feature matrix and the hidden feature matrix of graph. \\ 
$\textbf{x}_{\textit{v}}$, $\textbf{h}_{\textit{v}}$       & The feature vector and the hidden feature vector of the vertex \textit{v}, $\textbf{x}_{\textit{v}}$ $\in$ $\textbf{R}^{\textit{d}}$, $\textbf{h}_{\textit{v}}$ $\in$ $\textbf{R}^{\textit{d}}$. \\
\textit{L}, \textit{l}                                     & The number of GNN layers, and the index of each layer. \\
\hline
\end{tabular}
\vspace{-0.4cm}  
\end{table*}

\subsection{Graph Representation and Storage Format}

As illustrated in Table \ref{table:notation}, a graph is represented as \textit{G} = (\textit{V}, \textit{E}), where the set of vertices \textit{V} = \{\textit{$v_{0}$},\textit{$v_{1}$},$\dots$,\textit{$v_{n-1}$}\}, and the set of edges \textit{E} = \{$\textit{$e_{0}$},\textit{$e_{1}$},$\dots$,\textit{$e_{m-1}$}$\}. $\textit{$v_{i}$}$ $\in$ $\textit{V}$ denotes a vertex in the graph, and $\textit{$e_{ij}$}$ = ($\textit{$v_{i}$}$, $\textit{$v_{j}$}$) $\in$ $\textbf{E}$ denotes an edge pointing from $\textit{$v_{i}$}$ (source) to $\textit{$v_{j}$}$ (destination). The vertex relationship is represented by an adjacency matrix \textbf{A} with size \textit{n$\times$n}. If \textit{$e_{ij}$} $\in$ \textit{E}, then \textit{$A_{ij}$} = 1, otherwise \textit{$A_{ij}$} = 0. The degree matrix \textbf{D} is a diagonal matrix, where \textit{$D_{ii}$} represents the degree of vertex \textit{i}. The neighbourhood set of vertex \textit{v} is defined as \textit{N}(\textit{v}) = \{\textit{u} $\in$ \textit{V}$\mid$(\textit{v}, \textit{u}) $\in$ \textit{E}\}. Graph vertex attributes are represented as a multidimensional feature matrix \textbf{X} $\in$ $\textbf{R}^{\textit{n$\times$d}}$. $\textbf{x}_{\textit{v}}$ = ($\textit{$x_{0}$},\textit{$x_{1}$},$\dots$,\textit{$x_{d-1}$}$) is the feature vector of the vertex \textit{v}. For GNN, $\textbf{h}_{\textit{v}}^{(\textit{l})}$ represents the hidden feature vector of layer \textit{l} of vertex \textit{v}, $\textbf{W}_{\textit{v}}^{(\textit{l})}$ and $\textbf{W}_{\textit{E}}^{(\textit{l})}$ represent the feature matrix of layer \textit{l}.

\begin{figure}[h]
    \centering
    \vspace{-0.35cm}  
    \setlength{\abovecaptionskip}{0.2cm}     
    \setlength{\belowcaptionskip}{-0.5cm}   
    \includegraphics[width=1\textwidth]{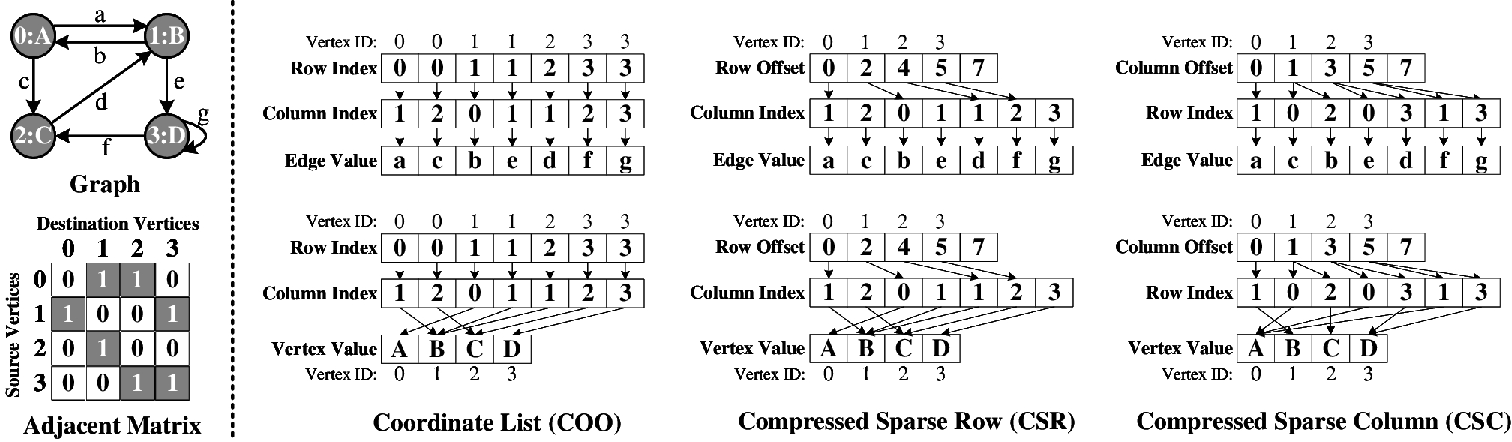}
    \caption{Storage format of adjacency matrix: The value includes the weights and features associated with vertices or edges. In TGC, one-dimensional features are typically employed, while GNNs commonly utilize high-dimensional feature vectors.}
    \label{fig:StorageFormat}
\end{figure}

There are three main storage formats for representing the non-zero elements in the matrix: coordinate list (COO), compressed sparse row (CSR), and compressed sparse column (CSC). As depicted in Figure \ref{fig:StorageFormat}, the COO format stores the row index, column index, and corresponding value (e.g., features, weights) of each non-zero element in separate arrays. Although simple and intuitive, the COO format contains redundant elements. To address this, the CSR and CSC formats were introduced to compress the redundant data. In the CSR format, the non-zero elements are stored in three arrays using row compression. The offset array stores the position of the first non-zero element in each row within the column array and the total number of non-zero elements. Similarly, the CSC format employs column compression. 

\subsection{Graph Processing Algorithms}

In this subsection, we will introduce two types of commonly used graph processing algorithms: TGC algorithms and GNN algorithms. We will explore typical examples and programming models related to each algorithm type. These programming models offer abstractions and guidelines for developing efficient graph processing algorithms. By providing this comprehensive introduction, our aim is to offer a clearer understanding of the execution process of these graph processing algorithms, enabling better acceleration and optimization strategies.

\subsubsection{Traditional Graph Computing (TGC) Algorithms}

The execution of such algorithms involves iterative updates of vertex information. Based on whether each iteration traverses all vertices, TGC algorithms can be classified into two categories: stationary graph algorithms and non-stationary graph algorithms \cite{2013Mizan, 2018SurveyScalable}. The stationary graph algorithms update all vertices in each iteration and rely on the current state of all vertices to compute the next state. Typical algorithms include PR, diameter estimation (DE), random walk with restart (RWR), etc. The non-stationary graph algorithms optimize the update process by considering vertex dependencies and update only the necessary portions of the graph. Typical algorithms include BFS, depth-first search (DFS), single-source shortest path (SSSP), etc.

\begin{figure}[ht]
    \centering
    \vspace{-0.4cm}  
    \setlength{\abovecaptionskip}{0.2cm}     
    \setlength{\belowcaptionskip}{-0.5cm}   
    \includegraphics[width=1\textwidth]{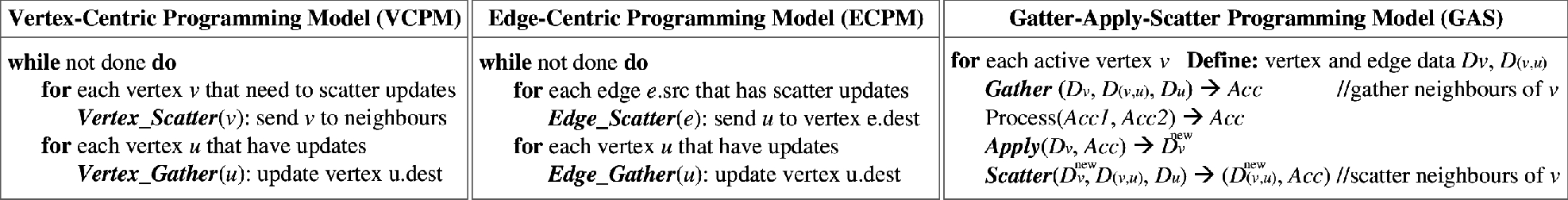}
    \caption{Programming mode of VCPM, ECPM \cite{2013X-stream} and GAS \cite{2012powergraph}: VCPM starts from each source vertex or active vertex, transmitting information to neighbors. ECPM starts from each edge, collecting information from source to destination vertices. GAS iterates through three stages, where Gather collects neighbor information from incoming edges, Apply updates central vertex attributes based on the collected information, and Scatter updates the neighbor vertex information of outgoing edges with the new value.}
    \label{fig:VCPM_ECPM_GAS}
\end{figure}

The programming models for TGC play a crucial role in efficiently processing large-scale graphs. As depicted in Figure \ref{fig:VCPM_ECPM_GAS}, three common TGC programming models are the vertex-centric programming model (VCPM) \cite{2010Pregel}, the edge-centric programming model (ECPM) \cite{2013X-stream}, and the gather-apply-scatter programming model (GAS) \cite{2012powergraph}. VCPM enables parallel execution of vertex processing and is commonly used in parallel systems. ECPM ensures continuous access to edge data, enhancing spatial locality. GAS offers flexibility, finding widespread usage in various software frameworks and accelerator designs. Developers can leverage the strengths of each approach to efficiently process graph data based on algorithm characteristics, underlying computing infrastructure and scenaro requirements.

\subsubsection{Graph Neural Network (GNN) Algorithms}

GNNs are a powerful class of models that integrate graph structures and NNs to effectively model and learn attributes and connections between vertices. GNN algorithms process input information iteratively layer by layer, to calculate information relevant to specific vertices. Several GNN models have been developed for diverse scenarios and can be categorized into four groups: convolutional graph neural network (ConvGNN), recurrent graph neural network (RecGNN), graph auto-encoder (GAE), and spatial-temporal graph neural network (STGNN) \cite{Pan2020comprehensive}. Among these, two ConvGNN models stand out as particularly prevalent and widely used: graph convolutional neural network (GCN) \cite{2017GCN} and graph attention network (GAT) \cite{2017GAT}. Currently, the majority of mainstream frameworks and accelerators are specifically optimized to enhance the performance of GCN and GAT.

A GCN can be generally abstracted in Equations \ref{equation:aggregate} \& \ref{equation:update}. During the aggregate stage, the GCN traverses the entire graph to collect and aggregate neighbor information for each vertex, resulting in an aggregated intermediate feature for the current layer $\textbf{h}_{\textit{N(v)}}^{\textit{(l)}}$, which is called hidden feature. In the update stage, the intermediate feature, along with the output feature from the previous layer $\textbf{h}_{\textit{N(v)}}^{\textit{(l)}}$, is combined to update the output feature vector of the current layer $\textbf{h}_{\textit{v}}^{\textit{(l)}}$. 

\begin{center}
\vspace{-0.5cm}  
\begin{equation}
\label{equation:aggregate}
\textbf{h}_{\textit{N}(\textit{v})}^{(\textit{l})} = \textbf{Aggregate}^{(\textit{l})}(\{ \textbf{h}_{\textit{u}}^{(\textit{l-1})} : \textit{u} \in \textit{N}(\textit{v}) \})
\end{equation}
\vspace{-0.5cm}  
\end{center}

\begin{center}
\vspace{-0.5cm}  
\begin{equation}
\label{equation:update}
\textbf{h}_{\textit{v}}^{(\textit{l})} = \textbf{Update}^{(\textit{l})}(\{\textbf{h}_{\textit{v}}^{(\textit{l-1})}, \textbf{h}_{\textit{N}(\textit{v})}^{(\textit{l})}\})
\end{equation}
\vspace{-0.6cm}  
\end{center}

Similar to a NN, a GNN consists of two parts: training and inference. During inference, forward propagation takes place, where aggregate and update processes alternate until reaching the maximum number of execution layers $\textit{L}$. In the training process, forward and backward propagation are iteratively performed to adjust the training parameters towards minimizing the loss function $\mathcal{L}$. The loss function $\mathcal{L}$ is utilized to measure the discrepancy between the predicted value and the true value. Generally, aggregate and update can be achieved by matrix multiplication, including SPMM or sparse-dense matrix multiplication (SPDMM). Here we give the hidden layer calculation formulas of GCN and GAT.

\begin{center}
\vspace{-0.8cm}  
\begin{equation}
\label{equation:GCN}
\textbf{H}^{\textit{l}} = \sigma \left( \widetilde{\textbf{D}}^{-\frac{1}{2}}\widetilde{\textbf{A}}\widetilde{\textbf{D}}^{-\frac{1}{2}}\textbf{H}^{\textit{l}-1}\textbf{W}^{\textit{l}-1} \right)
\end{equation}
\vspace{-0.5cm}  
\end{center}

\begin{center}
\vspace{-0.5cm}  
\begin{equation}
\label{equation:GAT}
\textbf{h}_\textit{v}^{(\textit{l})}=\sigma\left(\sum_{\textit{u}\in \textit{N}(\textit{v})}\alpha_\textit{vu}^{(\textit{l})}\textbf{W}^{(\textit{l-1})}\textbf{h}_\textit{u}^{(\textit{l-1})}\right)
\end{equation}
\vspace{-0.4cm}  
\end{center}

 The hidden layer of GCN is defined by Equation \ref{equation:GCN}, starting with the input layer $\textbf{H}^{(0)}=\textbf{X}$ and aiming for the output layer $\textbf{H}^{(L)}=\textbf{Z}$. The weight matrix, denoted as \textbf{W}, is trainable. During the aggregate stage, the feature matrix \textbf{H} is multiplied by the adjacency matrix \textbf{A}. In the update stage, the aggregated result is multiplied by the weight matrix \textbf{W} and then passed through a non-linear activation function $\sigma$. In contrast, GAT introduces attention mechanisms to learn the relative weights between connected vertices. The graph convolutional operation in GAT is defined by Equation \ref{equation:GAT}, where the attention weight $\alpha_\textit{vu}^{(\textit{l})}$ measures the connectivity strength between vertex \textit{v} and its neighbor \textit{u}.

%% file: tex/challenge.tex
In this section, we explore hardware-level challenges in graph processing execution. By exploring these challenges, we emphasize the crucial role of GPP methods in enhancing graph processing performance. We begin with a graph characteristics overview, detailing distinct execution behaviors that arise. We then analyze challenges stemming from these behaviors, emphasizing the significance of GPP in addressing them for efficient graph processing.

\subsection{Characteristics of Graph}

Real-world graphs usually have the following three characters:

$\bullet$ \textbf{Irregularity}: Due to the inherent uncertainty of real-world scenarios, connections between objects in graphs are highly random, resulting in unstable numbers of links and variable linked objects for each vertex. Unlike regular grid structures or linear arrangements found in images or text, graph topology exhibits significant irregularity.

$\bullet$ \textbf{Power-law Distribution}: Real-world graphs often exhibit a power-law distribution in vertex degrees \cite{2012powergraph}. This means most vertices have relatively few neighbors, while a few vertices have a significant number of neighbors. For example, in social networks, a small number of celebrities may have a substantial number of followers, while the majority of ordinary users have considerably fewer followers.

$\bullet$ \textbf{Large Scale}: Real-world graphs are often characterized by their immense size. As of 2023, Facebook \footnote{https://datareportal.com/essential-facebook-stats} boasts 2.25 billion vertices representing active users, while Twitter has 372.9 million vertices dedicated to active users. Furthermore, the World Wide Web \footnote{https://www.worldwidewebsize.com/index.php?lang=NL} encompasses a staggering minimum of 6.19 billion vertices, representing pages, along with hundreds of billions of edges to signify links. Additionally, real graphs frequently involve feature vectors with dimensions exceeding thousands, further amplifying the overall scale of the graph.

\subsection{Execution Behavior of Graph Processing}

\begin{figure}[h]
    \centering
    \vspace{-0.4cm}  
    \setlength{\abovecaptionskip}{0.2cm}     
    \setlength{\belowcaptionskip}{-0.5cm}   
    \includegraphics[width=0.8\textwidth]{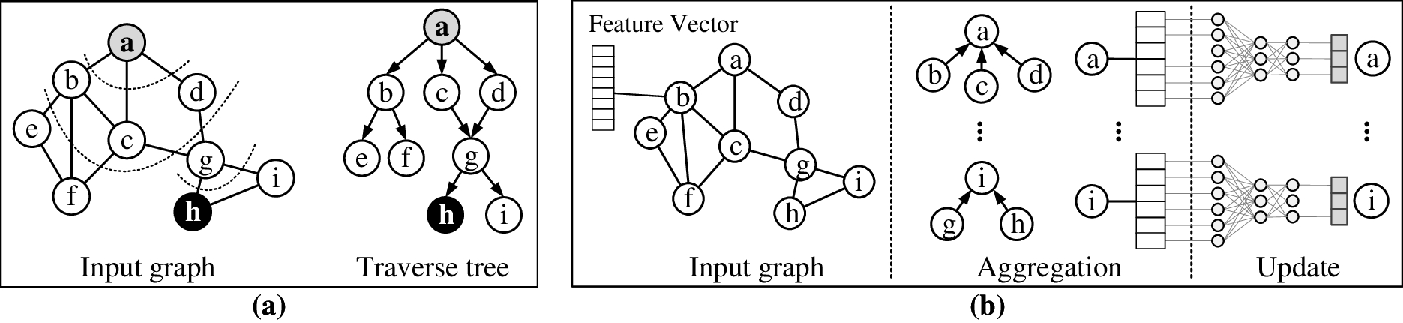}
    \caption{Examples of the execution process of TGC and GNN: (a) Traversal process of BFS; (b) Each iteration of GCN first performs an aggregate operation to gather vertex information and then performs a combine operation for NN transformation.}
    \label{fig:bfs_gcn}
\end{figure}

The execution behavior of graph processing algorithm is profoundly influenced by the characteristics of the graph data, data storage format and the algorithm model. Figure \ref{fig:bfs_gcn} visually compares BFS and GCN execution. To comprehensively analyze the execution behavior of graph processing, we focus on four aspects: computing mode, computing intensity, memory access mode, and data reuse rate. These aspects significantly impact algorithm efficiency and performance. Table \ref{table:behavior} provides an overview of the execution behavior of graph processing algorithms.

\begin{table*}[h]
\vspace{-0.3cm}  
\caption{Execution behaviors of graph processing}
\vspace{-0.3cm}  
\label{table:behavior}
\centering
\small
\begin{tabular}{|c|c|c|c|}
\hline
\multirow{2}{*}{\textbf{Behavior}}  & \multirow{2}{*}{\makecell{\textbf{Traditional Graph Computing}\\ \textbf{(TGC)} } } & \multicolumn{2}{c|}{\textbf{Graph Neural Network (GNN)}}       \\ \cline{3-4}
                                    &                                               & \multicolumn{1}{c|}{\textbf{Aggregate}}          & \textbf{Update}      \\ \hline
Computing Mode                      &  Dynamic \&  Irregular                        & \multicolumn{1}{c|}{Dynamic \&  Irregular}       & Static \&  Regular   \\ \hline
Computing Intensity                 &  Low                                          & \multicolumn{1}{c|}{Low}                         & High                 \\ \hline
Memory Access Mode                  &  Indirect \& Irregular                        & \multicolumn{1}{c|}{Indirect \& Irregular}       & Direct \& Regular    \\ \hline
Data Reuse Rate                     &  Low                                          & \multicolumn{1}{c|}{Low}                         & High                 \\ \hline
\end{tabular}
\vspace{-0.3cm}  
\end{table*}

$\bullet$ \textbf{Computing Mode}: TGC often involves traversing the neighbors of each vertex, a process also present in the aggregate stage of GNN. During traversal, information is passed between neighboring vertices, and the computation may vary depending on the graph structure and vertex attributes, resulting in a dynamic computing pattern. Additionally, the irregularity of the graph topology leads to a random number of neighbors being visited each time, creating an irregular computing pattern. On the other hand, the update stage processes the vertex attributes of entire graph at once, applying NN transformations to generate the output vertex feature vectors. As there is no need to traverse the graph iteratively, the computation is independent of input data, which means the computation is static. Furthermore, different vertices share the same neurons, resulting in a regular computing pattern in this stage.

$\bullet$ \textbf{Computing Intensity}: Computing intensity gauges the proportion of computation against data accessed from memory, which is a vital metric that reveals performance bottlenecks. In TGC, the vertex attributes typically comprise only a single element, yielding relatively simple operations like accumulation or comparison. As a result, the computing intensity is low and the performance is more constrained by memory access operations. Similarly, the aggregate phase in GNNs involves traversal operations to aggregate neighbors, exhibiting low computing intensity. Conversely, the update stage involves NN transformations to update high-dimensional vertex feature vectors for all vertices, exhibiting high computing intensity. So the performance in update stage is chiefly constrained by the computation itself.

$\bullet$ \textbf{Memory Access Mode}: Graph data is commonly stored in a compressed format. During graph traversal, the attribute data of neighboring vertices is accessed using the neighbor vertex numbers as indices, resulting in indirect memory access. The irregularity of the graph data leads to discontinuous addresses in this indirect memory access, resulting in irregular memory access patterns. In contrast, during the GNN update stage, all vertices undergo updates, and the high-dimensional vertex feature vectors are sequentially accessed directly from contiguous memory locations. Therefore, the GNN update stage exhibits a direct and regular memory access pattern.

$\bullet$ \textbf{Data Reuse Rate}: The resuse rate is a critical factor that affects computing efficiency and scalability during execution. It refers to the ability to reuse intermediate data, including intermediate results and cached input data. However, the irregularity of graph structures often leads to poor data locality and low data reuse rates in graph traversal processes between iterations. In contrast, the update stage of GNN demonstrates higher data reusability, as different vertices can share the same NN transformations and weights, leading to more efficient data utilization.

\subsection{Challenges of Graph Processing}

\begin{figure}[h]
    \centering
    \vspace{-0.2cm}  
    \setlength{\abovecaptionskip}{0.2cm}     
    \setlength{\belowcaptionskip}{-0.5cm}   
    \includegraphics[width=1\textwidth]{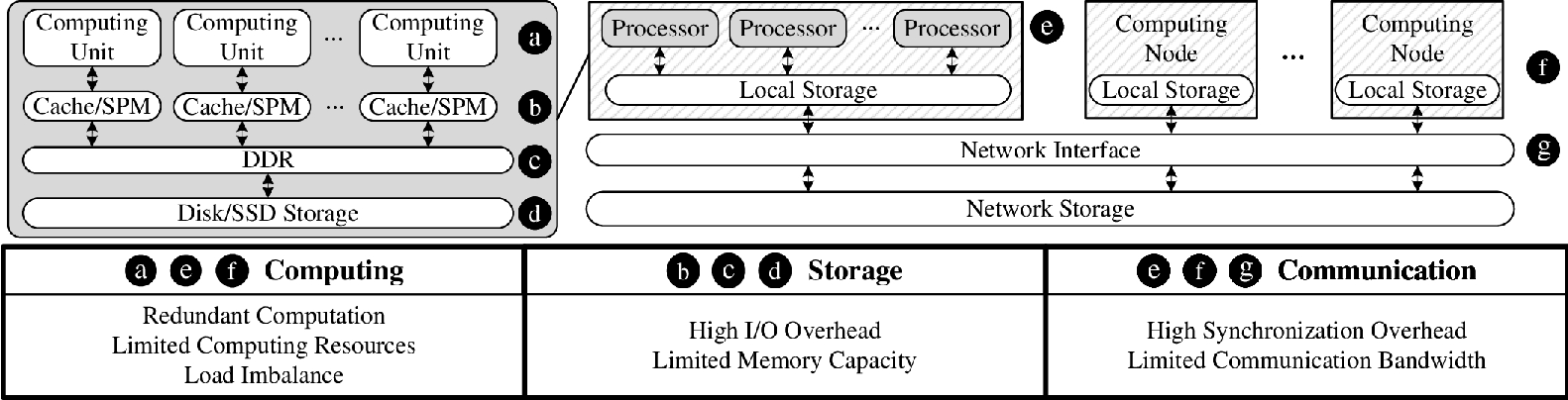}
    \caption{Hardware hierarchy and hardware-level challenges of graph processing: \protect\blackCircle{a} \protect\blackCircle{b} \protect\blackCircle{c} \protect\blackCircle{d} together form a processor \protect\blackCircle{e}, while \protect\blackCirclef{} represents a machine housing multiple processors. Also, \protect\blackCirclef{} signifies a computing node within a distributed system. The computing components of a parallel system encompass individual processors within a single machine and computing nodes within a distributed system. Typically, a machine housing multiple CPUs is referred to as a multi-CPUs system, and a distributed system with multiple CPUs is referred to as a CPU-clusters system.}
    \label{fig:hardware_hierarchy}
\end{figure}

Graph processing algorithms encounter various challenges due to their distinctive execution behavior. In this section, we will analyze these challenges from a hardware perspective, specifically considering computing, storage, and communication aspects. To provide a visual overview, Figure \ref{fig:hardware_hierarchy} depicts the general hardware hierarchy and summarizes the specific challenges faced by graph processing at the hardware level.

$\bullet$ \textbf{Challenges in Computing} \blackCircle{a} \blackCircle{e} \blackCirclef{}

\textit{Redundant Computation}: This occurs when the same information is repeatedly calculated during algorithm execution, leading to low utilization of computing resources and longer computing times, significantly impacting the performance and scalability of graph processing systems. The irregularity of the graph can cause multiple visits to the same vertex during traversal, resulting in redundant computations. Moreover, low data reusability during traversal means intermediate results are not effectively stored, requiring the recalculation of the same data. In parallel systems, different cs may redundantly process the same data when the graph is shared among multiple computing components.

\textit{Limited Computing Resources}: The problem arises when the size and complexity of the graph surpass the available computing power. Real-world graphs are often large, requiring massive computing power for graph algorithm execution. Additionally, some algorithms involve multiple iterations or recursive calculations, demanding a substantial amount of computing resources. For GNNs, the update stage involves computationally intensive calculations of high-dimensional feature vectors, potentially causing computing overload. 

\textit{Load Imbalance}: In large-scale graph processing, workload distribution across multiple components in a parallel system may become uneven. The irregular structure of the graph makes it challenging to partition vertices and edges into equal-sized subgraphs, leading to load imbalance. Some computing components become heavily loaded while others are underutilized, resulting in potential performance degradation. Addressing load imbalance is crucial for efficient and scalable graph processing.

$\bullet$ \textbf{Challenges in Storage} \blackCircle{b} \blackCircle{c} \blackCircle{d}

\textit{High I/O Overhead}: The I/O overhead in graph processing refers to the time and resources consumed during the movement of data between different storage layers. This overhead is mainly attributed to irregular data access patterns. Due to the irregularity of graphs, data is often not stored contiguously in memory, leading to a large number of random I/O operations, which are significantly slower than sequential access and result in high storage access overhead. The presence of high I/O overhead in graph processing can significantly impact overall performance and scalability of graph algorithms, leading to longer execution times and hindering the ability to efficiently process large graphs.

\textit{Limited Memory Capacity}: Large-scale graphs can exceed the available memory capacity, posing challenges to execution efficiency. Graph processing algorithms typically load the entire graph data into memory for efficient calculations. However, when the graph size surpasses the available memory, some parts of the graph must be swapped in and out of memory, leading to data overflow. This situation causes numerous off-chip memory accesses, resulting in additional delays and substantial performance degradation.

$\bullet$ \textbf{Challenges in Communication} \blackCircle{e} \blackCirclef{} \blackCircleg{}

\textit{High Synchronization Overhead}: Parallel graph processing requires coordination and synchronization among multiple computing components. As graph processing algorithms often depend on information from other components, frequent synchronization is necessary to ensure consistent computing results, resulting in high synchronization overhead. Moreover, inadequate local cache on components may lead to ineffective storage of shared data, resulting in repeated requests for the same data from other computing components, adding to the communication overhead. Excessive synchronization can cause communication delays and reduce the overall system execution efficiency.

\textit{Limited Communication Bandwidth}: This is another crucial problem in parallel graph processing, referring to situations where the available network bandwidth for communication between computing components is insufficient. In large-scale GNNs, high-dimensional feature vector data often demands significant data transmission, putting substantial pressure on the network bandwidth. Additionally, the irregular execution behavior of graph algorithms leads to unpredictable communication patterns, making it challenging to efficiently utilize the communication bandwidth. 

\subsection{The Significance of Graph Pre-processing (GPP)}


To address the challenges in graph processing, various graph processing frameworks and custom accelerators have been proposed. However, one of key factors in achieving efficient execution of graph algorithms is the pre-processing of graph data. GPP plays a vital role in optimizing the graph data and preparing it for efficient algorithm execution, thereby significantly improving the overall performance and scalability of graph processing systems.

Notably, the majority of graph processing work relies on GPP methods to tackle the complexities of large-scale graphs. A crucial step in processing such graphs is partition, which aims to achieve load balance and reduce communication overhead. Parallel frameworks like Giraph \cite{2011Giraph}, and AliGraph \cite{2019Aligraph} extensively utilize graph partition, as it significantly influences the data distribution across multiple computing components, enabling efficient parallel processing.

Moreover, GPP techniques are adopted to alleviate memory pressure and address the challenge of insufficient computing resources caused by large graphs. A widely adopted technique is the sampling-based mini-batch GNN training. GraphSage \cite{2017graphsage} is a classic sampling model used in various works, including GNN accelerators like HyGCN \cite{2020HyGCN}, AWB-GCN \cite{2020awbgcn}, and GNN training frameworks like NeuGraph \cite{2019Neugraph}, BGL \cite{2023BGL}, etc. Reorganization methods, such as those used in frameworks like CuSha \cite{2014cusha}, GridGraph \cite{2015Gridgraph}, and custom accelerators like FPGP \cite{2016fpgp}, HyGCN \cite{2020HyGCN}, etc., are also commonly employed to improve data access flow and further enhance the graph processing efficiency.

Furthermore, GPP becomes even more critical with the increase of heterogeneous platforms. GPP such as reorder and reconstruction, are usually performed on CPUs to prepare the graph data for acceleration on custom platforms. For example, GraphACT \cite{2020GraphACT} maximizes CPU and FPGA utilization, further optimizing the overall performance.

Additionally, technologies in NNs, such as sparsification and quantization, are widely extended to GNNs to reduce the amount of data and improve computational efficiency, typical works include DropEdge \cite{2020DropEdge}, BiFeat \cite{2022BiFeat}, etc.

In conclusion, effective GPP techniques are crucial in enhancing the performance, scalability, and efficiency of graph processing. They enable the optimization of graph data, distribution across computing components, and preparation of specialized accelerators, all of which play a crucial role in facilitating efficient graph processing. In the following sections, we propose our taxonomy and delve into a detailed analysis of existing GPP methods based on this taxonomy.

%% file: tex/taxonomy.tex
\begin{figure}[h]
    \centering
    \vspace{-0.3cm}  
    \setlength{\abovecaptionskip}{0.2cm}     
    \setlength{\belowcaptionskip}{-0.5cm}   
    \includegraphics[width=1\textwidth]{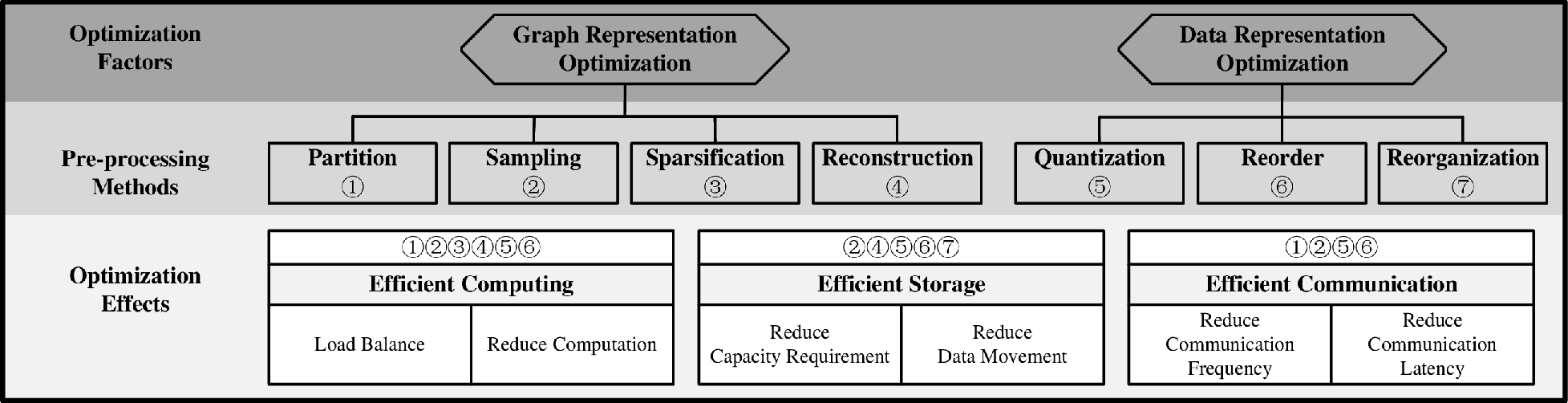}
    \caption{Taxonomy of GPP methods: The GPP methods are classified using a double-level decision framework. The first level considers optimization factors from an algorithmic perspective, while the second level examines optimization effects from a hardware perspective.}
    \label{fig:taxonomy}
\end{figure}

In this section, we present a comprehensive GPP methods taxonomy, utilizing a double-level decision framework, as shown in Figure \ref{fig:taxonomy}. \textit{In algorithmic perspective}, we categorize the seven methods into graph representation optimization and data representation optimization based on the optimization factors. \textit{In hardware perspective}, we analyze GPP effects, classifying GPP methods into efficient computing, storage, and communication. This framework enhances understanding. Next, we outline our taxonomy and explain the rationale behind our classification.

\subsection{Taxonomy in Algorithmic Perspective}

GPP entails two types of input graph data adjustments: graph representation optimization and data representation optimization. Graph representation optimization enhances graph algorithm performance by altering graph topology or density, while data representation optimization adjusts storage order or compresses data precision. By employing these two optimizations, researchers can explore various GPP approaches to improve the efficiency and effectiveness of graph algorithm execution while striking a balance between algorithm performance and accuracy.

$\bullet$ \textbf{Graph Representation Optimization}: This group encompasses partition, sampling, sparsification, and reconstruction techniques. Challenges in executing graph processing algorithms often arise from the irregularity of graph structures. These GPP methods modify input graph structures to enhance memory access and algorithm execution efficiency. Notably, partition, sampling, and sparsification may result in information loss, which could potentially affect the accuracy. In contrast, reconstruction solely alters the topology without impacting the algorithm's final outcomes.

$\bullet$ \textbf{Data Representation Optimization}: This group includes quantization, reorder, and reorganization techniques. Unlike graph representation optimization methods, these techniques do not alter the graph topology but instead focus on adjusting data storage. However, quantization reduces the precision of the data, introducing a trade-off between accuracy and execution efficiency. On the other hand, reorder and reorganization primarily adjust the data access mode to improve performance without affecting the algorithm's final outcome.

\subsection{Taxonomy in Hardware Perspective}

Considering the challenges highlighted in Section \ref{challenge} and the optimization goals, GPP offers opportunities to optimize graph algorithm execution in three aspects: efficient computing, efficient storage, and efficient communication. By effectively managing computing workloads, optimizing memory access, and minimizing communication overhead, GPP significantly improves the performance and resource utilization of graph processing systems.

$\bullet$ \textbf{Efficient Computing}: Improving computing efficiency can be achieved from two perspectives: load balance and computation reduction. Firstly, load balance ensures an even distribution of computational work among resources, maximizing their utilization and avoiding overloading computing resources. It aims to achieve efficient utilization of available computing units, leading to improved performance. Secondly, reducing computation overhead can be achieved by minimizing the computation amount, including the reduction of redundant computation and data volume. GPP methods for efficient computing include partition, sampling, sparsification, reconstruction, quantization and reorder.

$\bullet$ \textbf{Efficient Storage}: It can be achieved by reducing capacity requirements and reducing data movement. Firstly, to reduce capacity requirements, it involves minimizing the amount of data buffered on-chip, consequently reducing I/O overhead. An effective approach is to reduce the overall data volume. Secondly, minimizing data movement can significantly improve memory access bandwidth utilization. By effectively managing storage resources and optimizing data movement, the storage efficiency can be enhanced, then improving overall performance and resource utilization. GPP methods for efficient storage include sampling, reconstruction, quantization, reorder, and reorganization.

$\bullet$ \textbf{Efficient Communication}: It can be benefited from reducing communication frequency or latency. Firstly, by reducing communication frequency, the synchronization overhead of computing components can be minimized, including processors in a single machine and computing nodes in a distributed system. Improving data locality is a useful way to reduce data exchange needs between components. Secondly, reducing communication delay means making full use of communication bandwidth. One effective method is to reduce irregular and redundant communication requests. GPP methods for efficient communication include partition, sampling, quantization, and reorder.

%% file: tex/category-algorithm.tex
In this section, we will provide an overview of seven algorithmic GPP methods that are categorized into two distinct groups: graph representation optimization and data representation optimization. Our aim is to provide a comprehensive understanding of these methods by delving into their fundamental principles and exemplifying their typical algorithms. By the end of this section, readers will gain insights into the key concepts and practical implementation of these GPP methods. Typical algorithms of these GPP methods are listed in Table \ref{table:GPP-algorithm}.

\subsection{GPP Methods for Graph Representation Optimization}

\noindent{\ding{172} \textbf{Partition.}}
Graph partition is a critical process in parallel graph processing systems, serving as the pre-processing step in both TGC and GNN. It has two primary objectives. Firstly, it evenly distributes vertices and edges across different components to achieve load balance, which ensures an efficient utilization of computing resources. Secondly, it focuses on minimizing split edges to improve data locality, thereby reducing communication overhead between computing components. As a result of pursuing these two objectives simultaneously, graph partition is recognized as an NP-hard problem. The general expression for the graph partitioning process is as follows: 

\begin{center}
\vspace{-0.6cm}  
\begin{equation}
\label{equation:partition_1}
\textbf{Partition}(\textit{V})= \{{\bigcup}_{i} \textit{V}_\textit{i} \vert i=1,\cdots,\textit{k}; \forall i \neq j, \textit{V}_\textit{i} \cap \textit{V}_\textit{j} = \emptyset, \}
\end{equation}
\vspace{-0.5cm}  
\end{center}

\begin{center}
\vspace{-0.6cm}  
\begin{equation}
\label{equation:partition_2}
\{SubVertexSet_i \vert i=1,\cdots,k\} = \textbf{LoadBalance}(V_i, \cdots, V_k) 
\end{equation}
\vspace{-0.5cm}  
\end{center}

\begin{center}
\vspace{-0.6cm}  
\begin{equation}
\label{equation:partition_3}
\{Subgraph_i \vert i=1,\cdots,k\} = \textbf{MinEdgeCut}(E, \{SubVertexSet_i \vert i=1,\cdots,k\}) 
\end{equation}
\vspace{-0.6cm}  
\end{center}

Graph partition methods can be classified into two types \cite{2014BalanceEdgePartition} based on the cutting object: edge-cut partition and vertex-cut partition. In edge-cut partition, vertices are assigned to different subgraphs, resulting in edge-cuts that split edges crossing different subgraphs. In vertex-cut partition, edges are assigned to different subgraphs, and vertices may be replicated across multiple subgraphs. Partition can be performed offline, where the entire graph is divided in advance. For large-scale graph processing, streaming partition methods are used, continuously reading in edges or vertices for real-time partition. These online methods have lower time complexity and handle dynamic graphs efficiently. However, online partition may be less effective for lacking complete graphs. Next, we introduce some typical partition algorithms.

\begin{table*}[h]
\vspace{-0.3cm}  
\caption{GPP methods in algorithmic perspective}
\vspace{-0.3cm}  
\label{table:GPP-algorithm}
\centering
\small
\begin{tabular}{|c|p{11.2cm}|}
\hline
\textbf{Method} & \makecell[c]{\textbf{Works}}    \\
\hline
\makecell{Partition\\\ding{172}} & \makecell[l]{
                Edge-cut: RandomHash \cite{2010Pregel}, METIS \cite{1998METIS}, LDG\cite{2012LDG}, Fennel \cite{2014Fennel}, Chunk-based\cite{2016Gemini} \\
                Vertex-cut: GreedyEdge \cite{2012powergraph}, DBH \cite{2014DBH}, HDRF \cite{2015HDRF}, NE \cite{2017NE} } \\
\hline
\makecell{Sampling\\\ding{173}}  & \makecell[l]{
                Node-wise: GraphSAGE \cite{2017graphsage}, PinSage \cite{2018PinSage}, SSE \cite{2018SSE}, VR-GCN \cite{2019VR-GCN} \\
                Layer-wise: FastGCN \cite{2018FastGCN}, LADIES\cite{2019LADIES}, AS-GCN \cite{2021AS-GCN}  \\
                Subgraph-Based: GraphSAINT \cite{2019graphsaint}, Cluster-GCN \cite{2019Cluster-GCN}, RWT \cite{2021RWT}} \\    
\hline
\makecell{Sparsification\\\ding{174}}   & \makecell[l]{
                Accuracy-oriented: DropEdge \cite{2020DropEdge}, FastGAT \cite{2020FastGAT}, NeuralSparse \cite{2020NeuralSparse} \\
                Execution-oriented: GLT \cite{2021GLT}, Dynamic Pruning \cite{2021DyGNN}, AdaptiveGCN \cite{2021Adaptivegcn}} \\
\hline
\makecell{Reconstruction\\\ding{175}} &  \makecell[l]{
                Neighbor-merging: GraphACT \cite{2020GraphACT} \\
                Neighbor-caching: HAG \cite{2020Redundancy-free}}   \\
\hline
\makecell{Quantization\\\ding{176}} & \makecell[l]{
                Degree-free: EXACT \cite{2021EXACT}, Bi-GCN \cite{2021Bi-GCN}, DGCNN \cite{2021DGCNN} \\
                Degree-based: DegreeQuant\cite{2020DegreeQuant}, SGQuant \cite{2020SGQuant}, DBQ\cite{2022DBQ} } \\
\hline
\makecell{Reorder\\\ding{177}}  &  \makecell[l]{
                Matrix-based: RCM \cite{1984RCM}, ParallelRCM\cite{2014ParallelRCM} \\
                Degree-based: Hub-Sorting \cite{2017Cagra}, Hub-Clustering \cite{2018ReorderHC}, DBG \cite{2019DBGreorder}, GNNTiering \cite{2022GNNTiering} \\
                Community-based: NestedDissection \cite{2013mt-METIS}, LSH \cite{2021rubik}, Gorder \cite{2016Gorder}, Recall \cite{2017Recall}, \\ \ \ \ \ \ \ \ \ \ \ \ \ \ \ \ \ \ \ \ \ \ \ \ \ \ \ \ \ \ \ \ \ RabbitOrder \cite{2016RabbitOrder} }\\
\hline
\makecell{Reorganization\\\ding{178}}  & \makecell[l]{
                Shard: Interval-and-Shard \cite{2012graphchi, 2013X-stream, 2016fpgp, 2020HyGCN}, G-Shard \cite{2014cusha}\\ 
                Chunk-Block: GridGraph \cite{2015Gridgraph, 2017foregraph}, NXGraph \cite{2016NXgraph}, NeuGraph \cite{2019Neugraph} }  \\
\hline
\end{tabular}
\vspace{-0.3cm}  
\end{table*}

\begin{figure}[ht]
    \centering
    \vspace{-0.3cm}  
    \setlength{\abovecaptionskip}{0.2cm}     
    \setlength{\belowcaptionskip}{-0.3cm}   
    \includegraphics[width=0.75\textwidth]{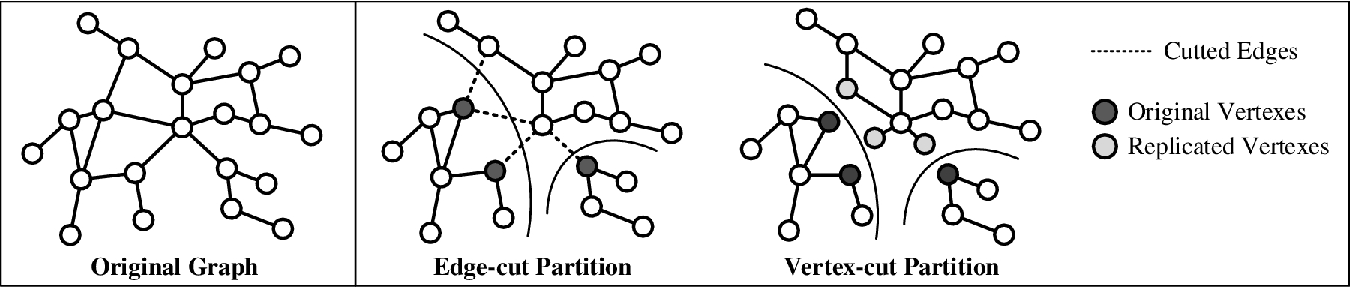}
    \caption{Graph partition methods: Edge-cut partition vs. vertex-cut partition}
    \label{fig:partition}
\end{figure}

$\bullet$ \textit{Edge-cut Partition}

\textit{RandomHash} \cite{2010Pregel} method is a simple heuristic approach commonly used as a baseline for initial graph partition experiments. It utilizes a random hashing function to partition the entire graph data. Although easy to implement and beneficial for small to medium-sized graphs due to its load balance and minimal storage requirements, it may not maintain locality within subgraphs for larger or more complex partition tasks, resulting in increased cut edges.

\textit{METIS} \cite{1998METIS}, a popular open-source multilevel graph partition method, follows three steps: coarsening, partitioning, and uncoarsening. Coarsening reduces the graph by merging vertices into supernodes; partitioning allocates vertices across partitions; uncoarsening projects the partitioning back to the original graph. However, due to the need to traverse the entire graph, METIS consumes significant memory and is less efficient for large-scale graphs.

\textit{LDG (Label Driven Greedy)} \cite{2012LDG} is a widely used streaming graph partition algorithm. It follows a multilevel approach, employing a label-driven greedy strategy and thinning to achieve balanced partitions with reduced edge cuts. This algorithm is highly efficient and effective in handling large-scale graphs with irregular structures.

\textit{Fennel} \cite{2014Fennel} is a streaming graph partition algorithm that follows a similar multilevel approach with LDG. It stands out in its initial partitioning stage, leveraging spectral partition techniques, offering better communication efficiency.

\textit{Chunk-based} partition \cite{2016Gemini} divides vertices into chunks and assigns them to cluster nodes. This method aims to achieve load balance by dynamically reassigning graph chunks, thus facilitating efficient parallel computing.

$\bullet$ \textit{Vertex-cut Partition}

\textit{GreedyEdge} \cite{2012powergraph} is a heuristic streaming graph partition method, which often serves as a practical baseline. It iteratively assigns edges to partitions based on a greedy strategy. Though it offers simplicity for large-scale graph partition, it may not always achieve the globally optimal partition and can be sensitive to the initial partition.

\textit{DBH} \cite{2014DBH} is a streaming graph partition method that utilizes a dynamic hashing function to assign vertices to partitions. It prioritizes dividing high-degree vertices to prevent splitting low-degree vertex communities and minimize communication during graph updates. DBH is memory-efficient but requires exploring the degree of each vertex.

\textit{HDRF} \cite{2015HDRF} is a streaming partition method that combines hierarchical partition and degree-based randomized fusion. The hierarchical partition recursively divides the graph into smaller subgraphs, facilitating large-scale graph handling. Degree-based randomized fusion groups highly connected vertices to improve communication efficiency.

\textit{NE} \cite{2017NE} is a heuristic method that considers the locality of neighbors to minimize vertex-cut.

\noindent{\ding{173} \textbf{Sampling.}}
In GNN training, employing a full-batch strategy introduces two key issues: memory constraints when traversing large graphs and slow convergence due to infrequent updates. Mini-batch training offers a solution by loading a limited number of vertices in each iteration, relieving memory pressure and accelerating convergence. Sampling selects a vertex subset for mini-batch creation, ensuring efficient training and scalability for large graphs. The sampling-based training addresses challenges in full-batch training, providing improved memory usage and faster convergence.

The sampling process in GNNs can be either online or offline. In offline sampling, the GNN model uses a fixed set of samples (i.e., subgraphs) from the graph data throughout the entire training or inference process, reducing computing and memory overhead. This approach is suitable when the graph data remains static during training. Conversely, in online sampling, the GNN model dynamically selects new samples from the graph data at each iteration during training. This adaptability allows the model to handle large-scale graphs that cannot fit entirely in memory and is particularly useful for processing dynamic graphs or streaming data. The graph sampling process is formulated as follows:

\begin{center}
\vspace{-0.8cm}  
\begin{equation}
\label{equation:sample_1}
SN (v) = \textbf{Sampling}^{(l)}(N(v))
\end{equation}
\vspace{-0.5cm}  
\end{center}

\begin{center}
\vspace{-0.6cm}  
\begin{equation}
\label{equation:sample_2}
\textbf{h}_{\textit{N}(\textit{v})}^{(\textit{l})} = \textbf{Aggregate}^{(\textit{l})}(\{ \textbf{h}_{\textit{u}}^{(\textit{l-1})} : \textit{u} \in \textit{SN}(\textit{v}) \})
\end{equation}
\vspace{-0.5cm}  
\end{center}

\begin{center}
\vspace{-0.6cm}  
\begin{equation}
\label{equation:sample_3}
\textbf{h}_{\textit{v}}^{(\textit{l})} = \textbf{Update}^{(\textit{l})}(\{\textbf{h}_{\textit{v}}^{(\textit{l-1})}, \textbf{h}_{\textit{N}(\textit{v})}^{(\textit{l})}\})
\end{equation}
\vspace{-0.7cm}  
\end{center}

Graph sampling methods can be classified into three types: node-wise sampling, layer-wise sampling, and subgraph-based sampling \cite{2021liu-sampling-survey}. Node-wise sampling selects vertex neighbors as subgraphs, but the total neighbors grow exponentially with increasing layers. Layer-wise sampling overcomes this by sampling a fixed number of vertices in each layer, but it may lead to sparser subgraphs in deeper layers. Subgraph-based sampling samples vertex and edge sets starting from an initial vertex, providing better independence. Next, we introduce some typical sampling models.

\begin{figure}[ht]
    \centering
    \vspace{-0.3cm}  
    \setlength{\abovecaptionskip}{0.2cm}     
    \setlength{\belowcaptionskip}{-0.4cm}   
    \includegraphics[width=1\textwidth]{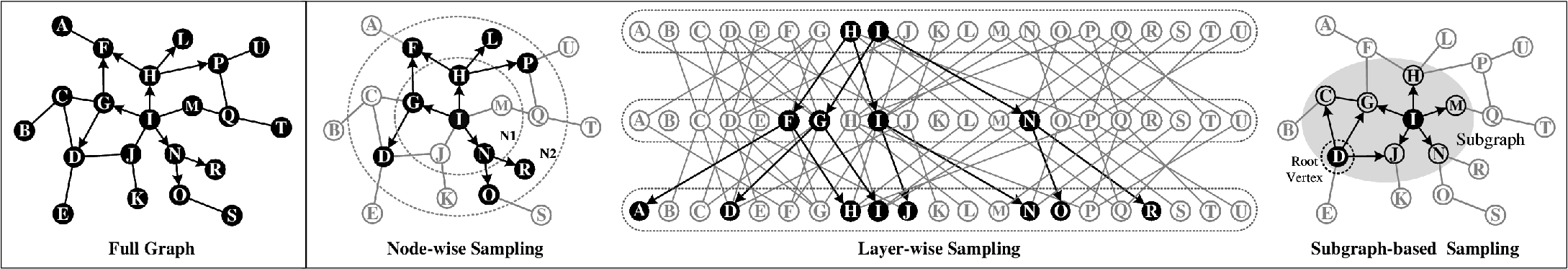}
    \caption{Graph sampling methods: Node-wise sampling vs. layer-wise sampling vs. subgraph-based sampling}
    \label{fig:sampling}
\end{figure}

$\bullet$ \textit{Node-wise Sampling}

\textit{GraphSAGE} \cite{2017graphsage} is a popular graph representation learning algorithm. It leverages local neighborhood information of vertices in a graph to perform inductive learning and learn node embeddings. During the training process, GraphSAGE samples a fixed-size neighborhood for each vertex, enabling efficient scalability to large graphs. While GraphSAGE primarily operates in an offline manner, it can also be combined with online sampling techniques for specific needs.

\textit{PinSAGE} \cite{2018PinSage} is an extension of GraphSAGE designed for personalized recommendations. It utilizes random walk for expansion and combines GraphSAGE neighborhood sampling and aggregation techniques with personalized information to enhance accuracy in personalized scenarios. Like GraphSAGE, PinSAGE is mainly used offline.

\textit{SSE} \cite{2018SSE} is an online sampling method used in large-scale graph processing. It performs random neighbor sampling for only 1-hop neighbors, avoiding the exponential growth of neighbors. SSE efficiently samples edges from the graph in real-time during processing, reducing memory requirements, and making it more scalable for processing large graphs.

$\bullet$ \textit{Layer-wise Sampling}

\textit{FastGCN} \cite{2018FastGCN} is an efficient and scalable graph sampling method proposed to accelerate GCN training. It uses a two-step strategy to select fixed-size neighborhoods for each vertex, improving computing efficiency. While primarily offline, it can be adapted for online sampling if needed.

\textit{LADIES} \cite{2019LADIES} is an online graph sampling method. It reduces computation overhead by discarding unnecessary vertices and edges during the sampling process, using a layer-wise discarding strategy for enhanced sampling efficiency.

\textit{AS-GCN} \cite{2021AS-GCN} is an online adaptive sampling method, selecting informative samples based on the importance. It samples a fixed number of vertices in each layer in a top-down manner, reusing vertices to improve training efficiency.

$\bullet$ \textit{Subgraph-based Sampling}

\textit{GraphSAINT} \cite{2019graphsaint} is an offline sampling method that estimates the probabilities of vertex and edge sampling separately. It aims to select informative and diverse subgraphs, thereby enhancing the training efficiency of GCNs.

\textit{Cluster-GCN} \cite{2019Cluster-GCN} is another offline graph sampling method that efficiently handles large graphs by employing clustering and mini-batch sampling. It significantly reduces memory usage during training, particularly for deep GCNs.

\textit{RWT} \cite{2021RWT} is an offline sampling method for training GCNs that utilizes random-walk sampling to construct mini-batches. It aims to sample diverse vertex neighborhoods, thereby enhancing the generalization capabilities of GCNs.

\noindent{\ding{174} \textbf{Sparsification.}}
GNNs use sparsification to enhance training and inference efficiency. This technique selectively removes unnecessary edges, reducing redundant computations and storage burdens. Additionally, we consider dynamic pruning as a sparsification method in this survey, as it creates a more compact graph for execution.

\begin{figure}[ht]
    \centering
    \vspace{-0.2cm}  
    \setlength{\abovecaptionskip}{0.2cm}     
    \setlength{\belowcaptionskip}{-0.3cm}   
    \includegraphics[width=1\textwidth]{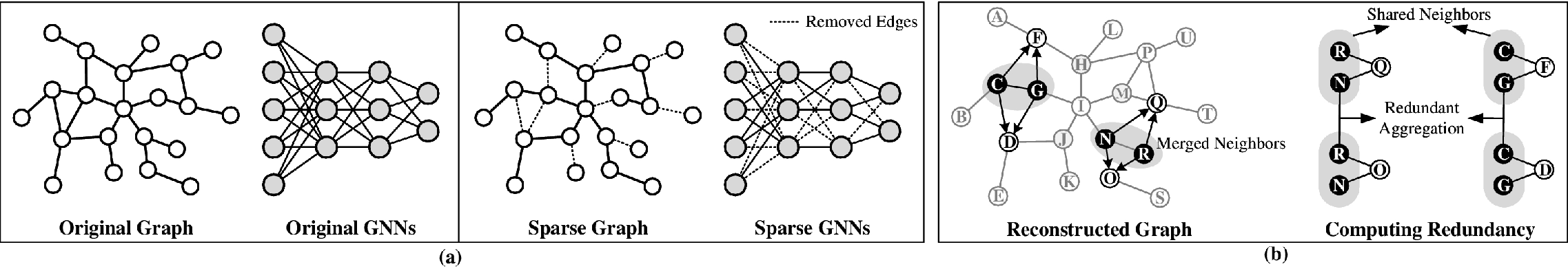}
    \caption{An illustration of sparsification and reconstruction process: (a) Graph sparsification through removing edges; (b) Graph reconstruction through merging shared neighbors}
    \label{fig:sparse-reconstruct}
\end{figure}

The sparsification process can be executed either online or offline, providing versatility for various applications. For graphs that evolve over time, sparsification efficiently removes redundant information and adapts to changing graph structures. In contrast, in static graphs, sparsification is valuable for reducing memory and computing overhead, especially for large graphs. The process of sparsification is shown in Figure \ref{fig:sparse-reconstruct}  (a), and it can be formulated as follows:

\begin{center}
\vspace{-0.5cm}  
\begin{equation}
\label{equation:sparse_1}
\textbf{A}_{sp} = Sparse(\textbf{A}) = RemoveEdge(\textbf{A}) \rightarrow Training\ or\ Inference\ GNN(\textbf{A}_{sp})
\end{equation}
\vspace{-0.7cm}  
\end{center}

Sparsification methods fall into two categories based on their purpose: accuracy-oriented and execution-oriented. Accuracy-oriented sparsification aims to enhance training accuracy by eliminating redundancies. However, it may lead to increased runtime due to more iterations. On the other hand, execution-oriented sparsification focuses on accelerating GNN execution and reducing computation and storage burdens. Next, we introduce some typical algorithms.

$\bullet$ \textit{Accuracy-oriented Sparsification}

\textit{DropEdge} \cite{2020DropEdge} is an offline method which randomly drops edges during training, sparsifying the graph and improving model robustness and generalization. It reduces computing complexity and memory requirements for deep GCNs.

\textit{FastGAT} \cite{2020FastGAT} is an offline method using effective resistance-based graph sparsification to reduce computing complexity while maintaining performance in GAT training.

\textit{NeuralSparse} \cite{2020NeuralSparse} is an offline method guided by a NN that prunes edges with low importance scores.

$\bullet$ \textit{Execution-oriented Sparsification}

\textit{GLT} \cite{2021GLT} generalizes the lottery ticket hypothesis from NNs to GNNs. By identifying a sparse subnetwork, GLT achieves comparable performance to the original dense network but with a smaller number of parameters. It is considered offline since it involves pre-training the dense network and then pruning it to find the winning ticket.

\textit{AdaptiveGCN} \cite{2021Adaptivegcn} introduces an adaptive sparsification technique for GCNs. It dynamically sparsifies the graph during training by adaptively removing edges based on their importance. This method can be categorized as online since it involves dynamic pruning of edges during training.

\textit{Dynamic Pruning} \cite{2021DyGNN} performs edge pruning during the training process to create a more compact and efficient graph representation. As it involves dynamic pruning during the training phase, it is considered as an online sparsification.

\noindent{\ding{175} \textbf{Reconstruction.}}
GNNs employ NN-based neighborhood message passing mechanisms to update target vertex representations by aggregating feature messages from neighboring source vertices. However, in complex graphs, a common scenario is the presence of multiple shared neighbors, leading to redundant calculations as these shared neighbors are aggregated repeatedly. To overcome this challenge, a technique called "reconstruction" is introduced. As shown in Figure \ref{fig:sparse-reconstruct} (b), this approach enables the reuse of intermediate aggregation results, thus enhancing the efficiency of computing and storage.

The reconstruction process causes no information loss since edges are not discarded. Reconstruction can be performed online or offline, depending on the applied scenarios. By adopting reconstruction, GNNs gain the promotion in terms of efficiency without compromising the accuracy of the graph data. The reconstruction technology can be categorized into two groups: neighbor-merging and neighbor-caching. Next, we introduce typical works in each groups.

$\bullet$ \textit{Neighbor-merging Reconstruction}

GraphACT \cite{2020GraphACT} merges shared neighbor pairs before training to effectively reduce redundant neighbor reduction operations. Since the merging occurs per epoch, it is performed online. Through data reconstruction, the data reuse rate is increased, resulting in improved execution efficiency.

$\bullet$ \textit{Neighbor-caching Reconstruction}

HAG \cite{2020Redundancy-free} caches and reuses intermediate results for subsequent vertices with similar neighborhoods. The caching of intermediate results occurs during the pre-processing step of the GNN, where the GNN traverses the graph and computes the intermediate aggregation results for each vertex. These results are then stored in a cache or lookup table. This combination of offline caching and online reuse reduces unnecessary overhead.

\subsection{GPP Methods for Data Representation Optimization}

\begin{figure}[ht]
    \centering
    \vspace{-0.2cm}  
    \setlength{\abovecaptionskip}{0.2cm}     
    \setlength{\belowcaptionskip}{-0.2cm}   
    \includegraphics[width=1\textwidth]{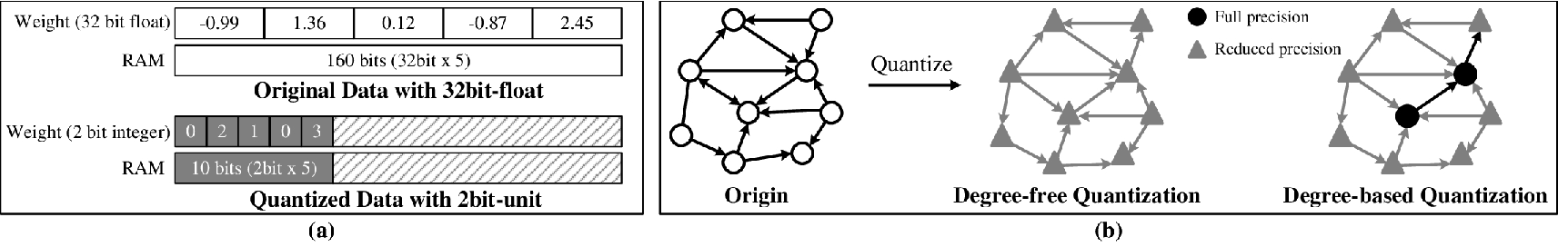}
    \caption{An illustration of quantization: (a) The 32-bit floating-point weights are quantized into 2-bit integer data; (b) Diagram of degree-free quantization and degree-based quantization.}
    \label{fig:quantization}
\end{figure}

\noindent{\ding{176} \textbf{Quantization.}}
Data quantization in GNNs involves converting continuous floating-point feature vectors into discrete values with a limited number of bits. This process reduces computational and memory overhead while sacrificing some model accuracy. Typically, high-dimensional feature data represented as 32-bit floating-point numbers is transformed into low-bit integer values, as shown in Figure \ref{fig:quantization} (a). Generalizing from NNs to GNNs, data quantization plays a crucial role in improving scalability and practicality on resource-constrained platforms.

Graph quantization can be divided into two categories: degree-free and degree-based, as shown in Figure \ref{fig:quantization} (b). Degree-free quantization directly extends classical NNs quantization methods to GNNs without considering vertex degrees. However, due to the irregular graph topology, methods that account for vertex degree are more suitable. Degree-based quantization methods address the irregular topology by quantizing based on vertex degrees, aiming to minimize data scale with an acceptable accuracy loss. Next, we will introduce some typical works.

$\bullet$ \textit{Degree-free Quantization}

\textit{EXACT} \cite{2021EXACT} introduces extreme activation compression for quantizing GNNs. By compressing training activations to a limited set of extreme values, storage and computation needs are minimized, enabling efficient large-scale graph training. The method employs quantization-aware training to mitigate accuracy degradation due to quantization.

\textit{Bi-GCN} \cite{2021Bi-GCN} proposes an offline quantization method that converts the continuous floating-point weights of GNNs into binary values. The approach also involves an adaptive scaling mechanism to preserve model accuracy. 

\textit{DGCNN} \cite{2021DGCNN} employs binary representations for vertex features and graph structures to achieve model compression. Vertex features are binary-coded, reducing memory overhead, while graph structures are encoded in binary forms.

$\bullet$ \textit{Degree-based Quantization}

\textit{DegreeQuant} \cite{2020DegreeQuant} proposes a quantization-aware training technique for GNNs, considering the irregular topology by incorporating vertex degree. This reduces computation and memory overheads while preserving quantization accuracy.

\textit{SGQuant} \cite{2020SGQuant} focuses on squeezing the last bit to achieve efficient training and inference.

\textit{DBQ} \cite{2022DBQ} identifies sensitive vertices in graph structures and applies a protective mask to ensure that these vertices perform full-precision calculations, while other vertices undergo quantization. This dynamic accuracy adjustment achieves greater acceleration without compromising classification accuracy.

\begin{figure}[ht]
    \centering
    \vspace{-0.3cm}  
    \setlength{\abovecaptionskip}{0.2cm}     
    \setlength{\belowcaptionskip}{-0.3cm}   
    \includegraphics[width=0.85\textwidth]{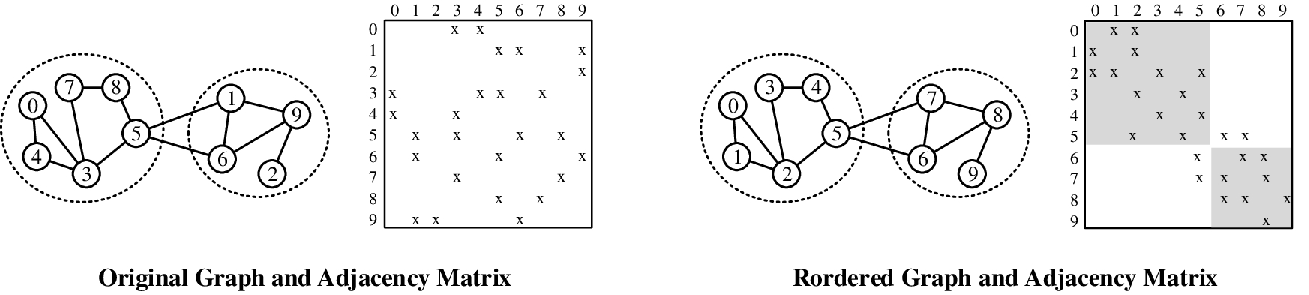}
    \caption{Graph reorder process: The data locality is improved compared to the original scattered matrix with unmodified IDs.}
    \label{fig:reorder}
\end{figure}

\noindent{\ding{177} \textbf{Reorder.}}
This method is a valuable data layout optimization technique that addresses the issue of poor memory locality in graph data caused by its structural irregularities. As depicted in Figure \ref{fig:reorder}, by relabeling vertex IDs, reorder improves data locality, taking reduced irregular memory accesses and improved memory access bandwidth utilization. Reorder can be applied to both TGC and GNN, enhancing the efficiency and scalability of large-scale graph processing.

Graph reorder techniques can be divided into three categories: matrix-based, degree-based, and community-based. Matrix-based reorder aims to compress the adjacency matrix bandwidth by reducing zero entries. While simple and fast, this approach may not fully consider the graph characters, leading to potential inefficiencies. In contrast, degree-based reorder takes vertex degrees into account. High-degree vertices are expected to be frequently accessed, and sorting them can improve cache utilization. Community-based reorder considers the power-law characteristics of the graph and groups closely connected vertices together, enhancing data locality. Next, we will describe some typical algoritms.

$\bullet$ \textit{Matrix-base Reorder}

\textit{RCM (Reverse Cuthill-McKee)} \cite{1984RCM} is a graph reorder algorithm that reduces matrix bandwidth in sparse graphs. It aims to improve cache utilization and memory access patterns by minimizing the distance between non-zero entries. The algorithm selects a starting vertex and performs BFS to find neighbors, ordering them based on their degrees. Papers \cite{2014ParallelRCM, 2017RCM} explore parallelization and distributed-memory implementation of RCM.

$\bullet$ \textit{Degree-base Reorder}

\textit{Hub-Sorting} \cite{2017Cagra} is a lightweight reorder technique that sorting high-degree vertices (hot vertices). It sorts hot vertices in descending order of degrees, placing them in a cacheline to enhance cache efficiency.

\textit{Hub-Clustering} \cite{2018ReorderHC} reorder technique clusters hot vertices with high-degree and places them at the beginning of the reordered sequence, but it does not further sort interior of these hot vertices clusters. Compared with Hub-Sorting, Hub-Clustering reduces the fine-grained reorder time overhead, but the spatial locality is relatively poor.

\textit{DBG (Degree-base Grouping)} \cite{2019DBGreorder} reorder adopts a coarse-grained sorting approach to preserve the graph structure while reducing the cache occupancy of hot vertices. It roughly divides the vertices into groups based on the degree and then arranges groups in descending order, while maintaining the original  vertex order within each group.

\textit{GNNTiering} \cite{2022GNNTiering} optimizes GNN training on GPUs using the Weighted Reverse Pagerank method. It stores high-scoring vertices with high outdegree in GPU memory, reducing large-scale data transmission over PCIe.

$\bullet$ \textit{Community-base Reorder}

\textit{NestedDissection} reorders and partitions graphs to reduce Cholesky factorization fill-in in sparse matrices. It recursively removes separator vertices and employs partition and reorder operations, with parallelization in mt-METIS \cite{2013mt-METIS}.

\textit{LSH} \cite{2021rubik} is a typical reorder technique that utilizes locality-sensitive hashing to group similar vertices together. 

\textit{Gorder} \cite{2016Gorder} is a heavyweight graph reorder algorithm that introduces a priority queue-based graph reorder algorithm (GO-PQ) to maintain frequently accessed vertices in a cacheline for improved cache utilization. While Gorder provides significant acceleration for large-scale graphs, it comes with high time overhead. Building on Gorder, \textit{ReCall} \cite{2017Recall} introduces a cache data reuse metric profile, yielding a heuristic pH, albeit with similarly high time overhead.

\textit{RabbitOrder} \cite{2016RabbitOrder} is the first just-in-time parallel reorder algorithm, which introduces a hierarchical communities-based approach derived from real-world hierarchical community structures. By mapping these communities into hierarchical caches, RabbitOrder leverages low-latency cache levels for improved performance.

\noindent{\ding{178} \textbf{Reorganization.}}
Large-scale graph processing faces challenges in caching all on-chip data, leading to substantial off-chip data movement. The irregular graph structure prompts random memory accesses, constraining performance. To tackle this, data reorganization enhances data flow by optimizing layout, bolstering reusability, and maximizing storage bandwidth. This technique splits the graph into slices, adjusting data order within each slice. Notably, reorganized slices are accessed sequentially, while partitioned subgraphs are processed concurrently. Next, we showcase typical works.

\begin{figure}[ht]
    \centering
    \vspace{-0.2cm}  
    \setlength{\abovecaptionskip}{0.2cm}     
    \setlength{\belowcaptionskip}{-0.5cm}   
    \includegraphics[width=1\textwidth]{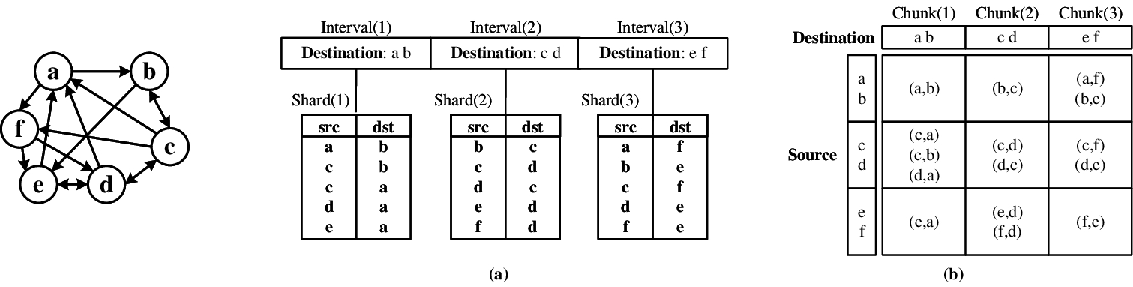}
    \caption{Examples of data reorganization: (a) Interval-and-Shard in Graphchi \cite{2012graphchi}; (b) Grid representation in GridGraph \cite{2015Gridgraph}}
    \label{fig:reorganization}
\end{figure}

\textit{Shard} technology, introduced in Graphchi \cite{2012graphchi}, optimizes data access on CPUs within a single machine. It splits graph data into intervals and shards (Figure \ref{fig:reorganization} (a)), organizing source vertices sequentially for efficient parallel sliding window-based data access. Shards find application in various works: X-stream offers a parallel implementation, CuSha tailors G-shards for GPUs, FPGP employs them on FPGAs, and HyGCN utilizes them for GNN inference \cite{2013X-stream, 2014cusha, 2016fpgp, 2020HyGCN}.

\textit{Chunk-Block} technique, from GridGraph \cite{2015Gridgraph}, divides the graph into grids. It splits vertices into P uniform chunks of fixed size and edges into PxP blocks (Figure \ref{fig:reorganization} (b)), utilizing dual sliding windows during computation to substantially reduce I/O overhead. Similar techniques are also employed in NXGraph \cite{2016NXgraph}, ForeGraph \cite{2017foregraph} and NeuGraph \cite{2019Neugraph}.

%% file: tex/category-hardware.tex
In the previous section, we provided an algorithmic perspective by introducing seven GPP methods. Now, our attention shifts to the hardware perspective, where we delve deeper into the optimization effects of these GPP methods. This section will be organized into three categories: \textit{efficient computing}, \textit{efficient storage}, and \textit{efficient communication}. We will explore practical implementations and examine related works that utilize these GPP methods to achieve optimization in each category. Our aim is to offer a comprehensive understanding of the hardware optimization possibilities that GPP methods present. Through the exploration of these related works, we will illustrate practical implementations and showcase the advancements made in this field.

\subsection{GPP Methods for Efficient Computing}

\begin{table*}[h]
\vspace{-0.4cm}  
\caption{Works with GPP methods for efficient computing}
\vspace{-0.3cm}  
\label{table:Efficient-computing}
\centering
\small
\begin{tabular}{|c|c|c|m{8cm}|}
\hline
\textbf{Effect}                                                               & \textbf{Method}            & \textbf{Algo.} & \multicolumn{1}{c|}{\textbf{Works}}  \\ \hline
\multirow{3}{*}{\begin{tabular}[c]{@{}c@{}}Load\\ Balance\end{tabular}}       & \multirow{3}{*}{Partition} & TGC   & Pregel \cite{2010Pregel}, Giraph \cite{2011Giraph}, Mizan \cite{2013Mizan}, PowerGraph \cite{2012powergraph}, GraphLab \cite{2012graphlab}, Gemini \cite{2016Gemini}, PowerLyra \cite{2019powerlyra} \\ \cline{3-4} 
                                                                              &                            & GNN   & AliGraph \cite{2019Aligraph}, DistDGL \cite{2020DistDGL}, PaGraph \cite{2020PaGraph}, Dorylus \cite{2021Dorylus}, QGTC \cite{2022QGTC} \\ \hline
\multirow{6}{*}{\begin{tabular}[c]{@{}c@{}}Reduce\\ Computation\end{tabular}} & Sampling                   & GNN   & AliGraph \cite{2019Aligraph}, DistDGL \cite{2020DistDGL}, AGL \cite{2020AGL}, PaGraph \cite{2020PaGraph}, 2PGraph \cite{2021-2PGraph}, P3 \cite{2021P3}, DGTP \cite{2022DGTP}, Pasca \cite{2022Pasca}, QGTC \cite{2022QGTC}, Sugar \cite{2023Sugar} \\ \cline{2-4} 
                                                                              & Sparsification             & GNN   & GLT \cite{2021GLT}, DyGNN \cite{2021DyGNN}  \\ \cline{2-4} 
                                                                              & Reconstruction             & GNN   & GraphACT \cite{2020GraphACT}, GCNInfer \cite{2020GcnInference}, HAG \cite{2020Redundancy-free}, I-GCN \cite{2021I-gcn}, ReGNN \cite{2022ReGNN} \\ \cline{2-4} 
                                                                              & Quantization               & GNN   & FPGAN \cite{2020FPGAN}, Bi-GCN \cite{2021Bi-GCN}, QGTC \cite{2022QGTC}\\\cline{2-4} 
                                                                              & Reorder                    & TGC   & Spara \cite{2020Spara}, GraphA \cite{2022GraphA}\\
\hline
\end{tabular}
\vspace{-0.6cm}  
\end{table*}

\subsubsection{Load Balance}
\label{Load Balance}

To achieve load balance in parallel graph processing systems with multiple computing components, graph partition is commonly utilized to evenly distribute vertices and edges among different components. Next, we will explore relevant works in graph partition.

$\bullet$ \textbf{Partition.}
Many parallel graph processing works utilize graph partition to achieve load balance, applicable to both TGC and GNN algorithms. Notably, a considerable number of parallel GNN training works employ a sampling-based strategy, thereby necessitating a thoughtful integration of graph partition with subsequent graph sampling.

\textit{Pregel} \cite{2010Pregel}, developed by Google, is a distributed TGC framework that operates within the batch synchronous parallel (BSP) model. It employs static graph partition to ensure an even distribution of computing tasks across computing nodes for load balance. \textit{Giraph} \cite{2011Giraph} extends Pregel's capabilities by integrating it with Hadoop. 

\textit{Mizan} \cite{2013Mizan} introduces dynamic partition to Pregel for dynamic workloads. Mizan dynamically adjusts partition boundaries, promoting dynamic load balance by redistributing workloads and maintaining vertex uniformity.

\textit{PowerGraph} \cite{2012powergraph} is a parallel TGC system, optimizing power-law graph processing through a greedy edge-cut partition strategy. It minimizes synchronization by only focusing on edge-connected vertices, mitigating the imbalance caused by intensive computations on a single node.

\textit{GraphLab} \cite{2012graphlab} is an open-source distributed TGC framework that employs a vertex-centric model and a variety of graph partition algorithms to guarantee a well-proportioned distribution of computational tasks.

\textit{PowerLyra} \cite{2019powerlyra} is an extension of GraphLab, achieving load balance by effectively distributing high-degree vertices across nodes. It introduces a customized hybrid partition strategy optimized for power-law graphs, utilizing an edge-cut partition for low-degree vertices and a vertex-cut partition for high-degree vertices.

\textit{Gemini} \cite{2016Gemini} is a distributed TGC framework, improving load balance through chunk-based graph partition. It dynamically allocates vertices to computing nodes, ensuring a fair distribution of workloads for better performance.

\textit{AliGraph} \cite{2019Aligraph} is Alibaba's distributed GNN training system that reduces edge crossings among threads via partition. The system offers a range of partition methods for selection and supports the addition of custom algorithms as plugins.

\textit{DistDGL} \cite{2020DistDGL} operates as a distributed GNN training framework, employing METIS for graph partition. It refines METIS by transforming its approach into a multi-constraint problem, leading to improved load balance.

\textit{PaGraph} \cite{2020PaGraph} supports GNN training using multiple GPUs within a single server, where each GPU is allocated an independent partition for training. It modifies the LDG partition method to suit graph sampling. To ensure even workload distribution across GPUs, every partition is equipped with a comparable count of training target vertices. 

\textit{Dorylus} \cite{2021Dorylus} is a distinctive distributed system for GNN training that leverages serverless computing for scalable and cost-effective solutions. It employs chunk-based graph partition to achieve load balancing effectively.

\textit{QGTC} \cite{2022QGTC} is a Tensor Core (TC)-based computing framework that can be used for GNN training and inference.  It employs METIS for graph partition, distributing distinct subgraphs to multiple GPUs for efficient computation.

\subsubsection{Reduce Computation}

As described in Section \ref{taxonomy}, there are two primary strategies to minimize computation: eliminating redundant computation and decreasing data volume. Effective GPP methods for eliminating redundancy include sparsification, reconstruction and reorder. Meanwhile, data volume can be reduced through GPP techniques such as sampling and quantization. Next, we will explore relevant works using these GPP methods.

$\bullet$ \textbf{Sparsification.}
By removing less influential or redundant edges, sparsification reduces the number of computations required during GNN operations, leading to faster training and inference times. 

\textit{UGS} \cite{2021GLT} presents a unified sparsification framework for GNN training, concurrently pruning both adjacency matrix and model weights. By applying the GLT sparsification method, it focuses on the most impactful vertices and edges, significantly diminishing the computational complexity of GNNs. Consequently, it leads to substantial savings in multiply accumulate (MAC) computation during SPMM operations.

\textit{DyGNN} \cite{2021DyGNN} accelerates GNN inference through a collaborative sparsification algorithm and customized architecture co-design. It addresses early vertex convergence and redundant neighbor aggregation, dynamic pruning to reduce graph density while preserving structure, thereby reducing computation. DyGNN incorporates a specialized Pruner for on-the-fly pruning and a streamlined pipeline design for rapid, low-latency inference.

$\bullet$ \textbf{Reconstruction.}
This method is employed in various GNN accelerators, effectively reducing aggregation redundancy. Some works use heterogeneous architectures, performing reconstruction on CPUs and graph processing on custom architectures. For further acceleration, more studies are customizing architectures for  real-time reconstruction.

\textit{GraphACT} \cite{2020GraphACT} proposes a CPU-FPGA heterogeneous architecture to expedite GNN training. It executes real-time reconstruction operations on the CPU, scanning each mini-batch to identify and merge neighbor pairs proactively. Reconstruction computes the highly reusable portions of the graph in advance, effectively precluding the need for subsequent redundant neighbor vertex reduction operations, significantly alleviating the computational burden.

\textit{GCNInfer} \cite{2020GcnInference} is a customized GNN inference accelerator, employing a reconstruction approach similar to GraphACT. Through offline preprocessing, it merges common neighbors and removes redundant edge computations for high-degree vertices. Notably, in the inference phase, which utilizes the entire graph, the effectiveness of redundancy elimination and computational enhancement surpasses that of the training phase with mini-batch subgraphs.

\textit{HAG} \cite{2020Redundancy-free} presents a novel hierarchical aggregation computation graph for managing intermediate results and reducing computation redundancy. Through online reconstruction, it caches intermediate aggregate results and utilizes them based on identified dependencies in subsequent iterations, effectively minimizing redundant computations during updates and enhancing overall computation speed for GNN training and inference.

\textit{I-GCN} \cite{2021I-gcn} is a GCN inference accelerator featuring an islandiation module for online reconstruction. This module clusters vertices to identify those with shared neighbors. During aggregate phase, common neighbor aggregation data is reused, eliminating redundant computations. Due to its custom architecture, I-GCN is superior in reconstruction speed.

\textit{ReGNN} \cite{2022ReGNN} designs a GNN accelerator that harmonizes algorithm and hardware through dynamic redundancy elimination. It presents a novel approach involving a neighborhood message passing algorithm, which pre-aggregates shared neighbors and stores reusable intermediate results. This method is supported by a dedicated architecture, effectively transforming redundancy elimination into performance enhancement.

$\bullet$ \textbf{Reorder.}
In ReRAM-based architectures, where computation occurs in memory, the presence of zero values in sparse matrices can result in considerable redundant computations. Reorder techniques enhance data locality, effectively mitigating sparsity and then improving overall energy efficiency.

\textit{Spara} \cite{2020Spara} is a ReRAM-based TGC accelerator. It employs the ReRAM crossbar size as a threshold for vertex reorder, optimizing workload density within the ReRAM crossbar. Therefore, it achieves a high computing energy efficiency.

\textit{GraphA} \cite{2022GraphA} is a TGC accelerator, comprising multiple ReRAM Graph Engines (RGEs). It assigns partitioned subgraphs to RGE units, then performs reorder based on the degrees, reducing sparsity to enhance computing resource utilization.

$\bullet$ \textbf{Sampling.}
By selecting a small subset of vertices and edges for training, graph sampling  decreases the vertex number involved in computing, ultimately reducing the overall computation complexity and facilitating faster iterations. Sampling technique is widely adopted, finding its application in various frameworks designed for GNN training. 

Many distributed GNN training frameworks integrate partition with sampling. Some frameworks start by partitioning the large graph into subgraphs, which are then allocated to individual computing nodes for sampling. Such frameworks include AliGraph \cite{2019Aligraph}, DistDGL \cite{2020DistDGL}, PaGraph \cite{2020PaGraph}, 2PGraph \cite{2021-2PGraph}, P3 \cite{2021P3}, DistDGLv2 \cite{2022DistDGLv2}, Pasca \cite{2022Pasca}, and Sugar \cite{2023Sugar}. In contrast, some frameworks directly sample the original graph and then distribute them to various computing nodes. Such frameworks include AGL \cite{2020AGL} and DGTP \cite{2022DGTP}. The former one performs sampling within the assigned subgraph independently at each compute node, however, it may result in information loss, affecting convergence. The latter one samples the entire graph, preserving information integrity but consuming more resources for sampling.

$\bullet$ \textbf{Quantization.}
By conversing high-precision floating-point data into low-precision integer data, quantization diminishes the overall workload, significantly curtailing the requirement for computing resources. Also, post quantization, matrix multiplication can be streamlined into simpler logical operations, reducing the computation complexity.

\textit{FPGAN} \cite{2020FPGAN} operates as an FPGA-based GAT inference accelerator. It quantizes input features, converting the original floating-point multiplication into a shift operation. Furthermore, its customized architecture enhances calculation speed while conserving computational resources.

\textit{Bi-GCN} \cite{2021Bi-GCN} introduces a binary GCN that binarizes both network parameters and vertex features. This conversion allows for the transformation of complex SPMM into straightforward binary operations, thus achieving acceleration.

\textit{QGTC} \cite{2022QGTC}, a TC-based GNN computing framework, proposes a novel quantized low-bit arithmetic design. This design utilizes low-bit data representation and bit-decomposed computation to convert complex high-precision floating-point matrix multiplication into faster bit-matrix multiplication, optimizing computing speed.

\subsection{GPP Methods for Efficient Storage}

\begin{table*}[h]
\vspace{-0.4cm}  
\caption{Works with GPP methods for efficient storage}
\vspace{-0.3cm}  
\label{table:Efficient-storage}
\centering
\small
\begin{tabular}{|c|c|c|m{8cm}|}
\hline
\textbf{Effect}                                                                          & \textbf{Method}                 & \textbf{Algo.}       & \multicolumn{1}{c|}{\textbf{Works}}     \\ \hline
\multirow{3}{*}{\begin{tabular}[c]{@{}c@{}}Reduce\\ Capacity\\ Requirement\end{tabular}} & \multirow{2}{*}{Sampling}       & \multirow{2}{*}{GNN} & AliGraph \cite{2019Aligraph}, DistDGL \cite{2020DistDGL}, AGL \cite{2020AGL}, PaGraph \cite{2020PaGraph}, 2PGraph \cite{2021-2PGraph}, \\ 
                                                                                         &                                 &                      & P3 \cite{2021P3}, DGTP \cite{2022DGTP}, Pasca \cite{2022Pasca}, QGTC \cite{2022QGTC}, Sugar \cite{2023Sugar} \\ \cline{2-4} 
                                                                                         & Quantization                    & GNN                  & SGQuant \cite{2020SGQuant}, EXACT \cite{2021EXACT}, DGCNN \cite{2021DGCNN}, BiFeat \cite{2022BiFeat}, DBQ \cite{2022DBQ} \\ \hline
\multirow{8}{*}{\begin{tabular}[c]{@{}c@{}}Reduce\\ Data\\ Movement\end{tabular}}        & Reconstruction                  & GNN                  & GraphACT \cite{2020GraphACT}, GCNInfer \cite{2020GcnInference}, HAG \cite{2020Redundancy-free}, I-GCN \cite{2021I-gcn}, ReGNN \cite{2022ReGNN} \\ \cline{2-4} 
                                                                                         & \multirow{2}{*}{Reorder}        & TGC                  & Cagra \cite{2017Cagra}  \\ \cline{3-4} 
                                                                                         &                                 & GNN                  & GCNInfer \cite{2020GcnInference}, Rubik \cite{2021rubik}, H-GCN \cite{2022H-GCN}, GNNTiering \cite{2022GNNTiering} \\ \cline{2-4} 
                                                                                         & \multirow{5}{*}{Reorganization} & \multirow{4}{*}{TGC} & GraphChi \cite{2012graphchi}, TurboGraph \cite{2013TurboGraph}, X-Stream \cite{2013X-stream}, CuSha \cite{2014cusha}, \\
                                                                                         &                                 &                      & FPGP \cite{2016fpgp},  GridGraph \cite{2015Gridgraph}, NXGraph \cite{2016NXgraph}, ForeGraph \cite{2017foregraph}, \\
                                                                                         &                                 &                      & FPGA-Cache \cite{2021FPGA-cache}, Graphicionado \cite{2016graphicionado}, GraphDynS \cite{2019GraphDynS},  \\
                                                                                         &                                 &                      & GraphR \cite{2018GraphR}, GraphSAR \cite{2019graphSAR} \\ \cline{3-4} 
                                                                                         &                                 & GNN                  & NeuGraph \cite{2019Neugraph}, HyGCN \cite{2020HyGCN} \\ \hline
\end{tabular}
\vspace{-0.4cm}  
\end{table*}

\subsubsection{Reduce Capacity Requirement}
In graph processing, optimizing on-chip storage for initial data and intermediate results is crucial, since insufficient on-chip memory leads to slow off-chip I/O. A viable optimization strategy is to reduce the overall data volume, and two effective techniques for achieving this in GPP are sampling and quantization.

$\bullet$ \textbf{Sampling.}
In contrast to full-batch GNN training, the sampling-based strategy only requires caching mini-batches, substantially reducing memory demands. As a result, this approach is more easily scalable and deployable, and it finds widespread usage in numerous studies. An overview of relevant distributed GNN training frameworks is presented in Table \ref{table:Efficient-storage}. Given that these frameworks have been introduced in Section \ref{Load Balance}, redundant explanations are omitted here.

$\bullet$ \textbf{Quantization.}
By employing fewer bits to represent each value, quantization significantly reduces overall memory consumption. This reduction in memory requirements is particularly advantageous when on-chip memory is limited.

\textit{SGQuant} \cite{2020SGQuant} employs a multi-granularity quantization approach, utilizing layer-wise, component-wise, and topology-aware quantization granularity. It compresses GNN features while minimizing accuracy loss. Experiments show that \textit{SGQuant} achieves a remarkable memory capacity reduction of up to 31.9 times while maintaining high training accuracy.

\textit{EXACT} \cite{2021EXACT} is a GPU-based GNN training solution specifically designed for deep GNNs. Addressing the challenge of substantial activation storage requirements during training, EXACT employs compression techniques to markedly reduce memory overhead. Experimental results indicate that EXACT achieves memory savings of up to 32 times.

\textit{DBQ} \cite{2022DBQ} introduces degree-based quantization, an efficient method to reduce memory usage. DBQ strategically quantifies vertices with minimal influence, striking a balance between precision loss and quantization compression ratio. This innovative approach effectively addresses memory constraints while maintaining a high level of accuracy.

\textit{DGCNN} \cite{2021DGCNN} stands out by utilizing binary representations for both vertex features and graph structures. This efficient approach involves the conversion of node features into binary codes, thereby reducing memory overhead. Moreover, the encoding of graph structures in binary format further contributes to computational efficiency and model compression.

\textit{BiFeat} \cite{2022BiFeat} puts forth a novel GNN feature quantization approach. This technique accommodates both scalar and vector quantization methods, effectively reducing the memory footprint of GNN feature data. While scalar quantization, based on the logarithmic method, offers quick GPP with modest compression, vector quantization achieves extensive data compression, making it suitable for large graphs albeit with more time-consuming compression calculations.

\subsubsection{Reduce Data Movement}
The irregularity of graph topology leads to inefficient and random memory access patterns, resulting in frequent data movement. Reorder and reorganization methods enhance dataflow access, thus improving memory bandwidth utilization. Additionally, the reconstruction technique consolidates redundant data, effectively reducing unwarranted transfers. Next we introduce works that leverages these GPP methods for optimization.

$\bullet$ \textbf{Reconstruction}
Reusing intermediate results through the reconstruction method effectively reduces redundant data movement. Notably, GraphACT \cite{2020GraphACT}, GCNInfer \cite{2020GcnInference}, and I-GCN \cite{2021I-gcn} strategically decrease data transfer volumes by pre-merging shared vertices. Additionally, HAG \cite{2020Redundancy-free} and ReGNN \cite{2022ReGNN} enhance storage efficiency by caching reusable intermediate outcomes, effectively mitigating the need for repeated data reading.

$\bullet$ \textbf{Reorder.}
This method significantly enhances data locality through the strategic rearrangement of graph vertices. This transformation effectively converts irregular memory access patterns into more regular sequences, thereby optimizing cache storage efficiency and mitigating cache misses.

\textit{Cagra} \cite{2017Cagra} serves as a cache-optimized memory graph framework, designed to expedite TGC. This framework employs hub-sorting for vertex reorder. Cagra focuses on enhancing the gather phase of pull-based graph algorithms, strategically aggregating frequently accessed vertices based on out-degree to bolster cache efficiency.

\textit{Rubik} \cite{2021rubik} is a GCN inference accelerator that employs LSH reorder to enhance data reuse during aggregation.

\textit{GCNInfer} \cite{2020GcnInference} is a GCN inference accelerator that executes GCN using SPMM. To enhance data locality and maximize bandwidth utilization, it employs the RCM algorithm for vertex reorder. 

\textit{H-GCN} \cite{2022H-GCN} stands as a hybrid accelerator utilizing PL (programmable logic) and AIE (AI engine) within the Xilinx Adaptive Computing Acceleration Platform for high-performance GNN inference. It utilizes open-source tool mt-metis to implement graph reorder, effectively clustering vertices with shared neighbors and optimizing data locality.

\textit{GNNTiering} \cite{2022GNNTiering} is a multi-GPU GNN training framework that uses the GraphSage model for mini-batch training. It strategically improves storage access efficiency by employing statistical techniques to identify data locality. GNNTiering uses reorder method to allocate high-degree vertices to the GPU cache and routes low-degree vertices to the CPU. This minimizes PCIe transmission of high-dimensional feature vectors between the CPU and GPU.

$\bullet$ \textbf{Reorganization.}
Large-scale graph processing often involves frequent data exchange between memory and off-chip storage due to memory limitations. Reorganization techniques offer a solution by optimizing data layout in advance, thereby enhancing data read and write efficiency while reducing the overhead associated with data movement.

\textit{GraphChi} \cite{2012graphchi} is a prominent framework for efficient large-scale TGC. It introduces the "Interval-and-Shard" technique, partitioning graphs into intervals and shards through offline GPP, enhancing memory usage and computation by promoting data locality. GraphChi employs the Parallel Sliding Window (PSW) method to sequentially access intervals and shards, minimizing random disk access, thereby improving storage bandwidth utilization and achieving high performance on a single machine. \textit{TurboGraph} \cite{2013TurboGraph} and \textit{X-stream} \cite{2013X-stream} present parallel improvements for GraphChi. This technique is also used in TGC accelerators Graphicionado \cite{2016graphicionado} and GraphDynS \cite{2019GraphDynS}.

\textit{CuSha} \cite{2014cusha} is a CUDA-based framework, accelerating TGC on a single machine with multiple GPUs. It introduces G-Shard and Concatenated Windows (CW) to harness GPU parallelism by splitting edges into shards and sequentially accessing storage using CW. This optimization ensures coalesced memory access, reducing data flow irregularities. 

\textit{FPGP} \cite{2016fpgp} stands as a FPGA-based streaming framework tailored for TGC. Similar to GraphChi, FPGP also adopts the Interval-and-Shard methodology. In this scheme, the graph's vertices are divided into $P$ intervals, while the edges are partitioned into $P^2$ sub-shards. Notably, updates are executed in sub-shard units. By leveraging Interval-and-Shard technology, FPGP optimizes storage access bandwidth utilization and achieves streaming execution of TGC algorithms.

\textit{GridGraph} \cite{2015Gridgraph} enables large-scale TGC on a single machine through 2D-partitioning and dual sliding windows (DSW). It splits graph into vertex chunks and edge blocks, utilizing DSW to streamline edge streaming and dynamic vertex updates, reducing storage I/O overhead. Notably, the GPP costs in GridGraph is lower than that in GraphChi.

\textit{NXGraph} \cite{2016NXgraph}, a single-machine TGC framework, employs a reorganization technique similar to GridGraph. Notably, it arranges destination vertices in descending order to enable sequential storage of edges with the same destination.

\textit{GraphR} \cite{2018GraphR} is the pioneering work that leverages ReRAM to expedite TGC. It employs a method akin to GridGraph, dividing blocks and generating ordered edge lists to enhance the flow of data access. A similar work is \textit{GraphSAR} \cite{2019graphSAR}.

\textit{ForeGraph} \cite{2017foregraph} is a multi-FPGA TGC accelerator. It divides the graph into Intervals and Blocks, subsequently breaks them down into Sub-Intervals and Sub-Blocks. ForeGraph adopts a destination-first replacement strategy to optimize off-chip I/O. Notably, this reorganization technique is also harnessed in the FPGA-based accelerator FPGA-Cache \cite{2021FPGA-cache}.

\textit{NeuGraph} \cite{2019Neugraph} is a scalable parallel GNN training framework that, employing chunk and block division for efficient processing like GridGraph. By processing edge blocks sequentially, NeuGraph minimizes random access inefficiencies. Its graph-aware dataflow engine intelligently breaks graph data into blocks, enabling parallel execution on GPUs.

\textit{HyGCN} \cite{2020HyGCN} is a classic GCN inference accelerator. It borrows the Interval-and-Shard concept from GraphChi to split graph data in GNNs, assigning vertices within each interval continuous numbers and storing them sequentially. This data reorganization facilitates the sequential caching of feature vectors, resulting in enhanced bandwidth utilization.

\subsection{GPP Methods for Efficient Communication}

\begin{table*}[h]
\vspace{-0.3cm}  
\caption{Works with GPP methods for efficient communication}
\vspace{-0.3cm}  
\label{table:Efficient-communication}
\centering
\small
\begin{tabular}{|c|c|c|m{7.8cm}|}
\hline
\textbf{Effect}                                                                             & \textbf{Method}           & \textbf{Algo.}       & \multicolumn{1}{c|}{\textbf{Works}} \\ \hline
\multirow{3}{*}{\begin{tabular}[c]{@{}c@{}}Reduce\\ Communication\\ Frequency\end{tabular}} & \multirow{3}{*}{Partition}& \multirow{2}{*}{TGC} & Pregel \cite{2010Pregel}, Giraph \cite{2011Giraph}, Mizan \cite{2013Mizan}, PowerGraph \cite{2012powergraph},  \\
                                                                                            &                           &                      & GraphLab \cite{2012graphlab}, Gemini \cite{2016Gemini}, PowerLyra \cite{2019powerlyra} \\ \cline{3-4} 
                                                                                            &                           & GNN                  & AliGraph \cite{2019Aligraph}, Dorylus \cite{2021Dorylus}, DistGNN \cite{2021DistGNN}, MultiGCN \cite{sungongjian2022MultiAccSys} \\ \hline
\multirow{5}{*}{\begin{tabular}[c]{@{}c@{}}Reduce\\ Communication\\ Latency\end{tabular}}   & Quantization              & GNN                  & BiFeat \cite{2022BiFeat}, QGTC \cite{2022QGTC} \\ \cline{2-4} 
                                                                                            & \multirow{3}{*}{Sampling} & \multirow{3}{*}{GNN} & AliGraph \cite{2019Aligraph}, DistDGL \cite{2020DistDGL}, AGL \cite{2020AGL}, PaGraph \cite{2020PaGraph},  \\ 
                                                                                            &                           &                      & P3 \cite{2021P3}, 2PGraph \cite{2021-2PGraph}, DistDGLv2 \cite{2022DistDGLv2}, DGTP \cite{2022DGTP}, \\ 
                                                                                            &                           &                      & Pasca \cite{2022Pasca}, Sugar \cite{2023Sugar} \\ \cline{2-4} 
                                                                                            & Reorder                   & TGC                  & RabbitOrder \cite{2016RabbitOrder} \\ 
                                                                                            \hline
\end{tabular}
\vspace{-0.4cm}  
\end{table*}

\subsubsection{Reduce Communication Frequency}

Within a parallel graph processing system, reducing the communication frequency between computing components is paramount to mitigating communication overhead. An impactful pre-processing strategy involves enhancing data locality through graph partitioning, consequently decreasing communication demands among computing nodes. The subsequent analysis delves into the mechanics of this approach.

$\bullet$ \textbf{Partition.}
To decrease communication frequency, effective graph partition must minimize edge cutting, reducing inter-subgraph data dependency. This objective has led to the development of various partition strategies prioritizing enhanced data locality. However, achieving a balance between load distribution and data locality often involves trade-offs. It's important to note that, for power-law graphs, edge-cut partition typically outperforms vertex-cut partition due to its ability to reduce inter-node communication by copying mirror vertices. Below, we introduce some typical works. Other works listed in Table \ref{table:Efficient-communication} are detailed in Section \ref{Load Balance}, and not be reiterated here.

\textit{DistGNN} \cite{2021DistGNN} is a distributed framework, performing GNN training on CPU-clusters. It uses DBH \cite{2014DBH} partition, a vertex-cut method to reduce communication frequency. 

\textit{MultiGCN} \cite{sungongjian2022MultiAccSys}, a multi-node GCN accelerator, introduces a round partition method, dividing replicas into processing rounds to store them entirely in on-chip, reducing off-chip data access. 

\subsubsection{Reduce Communication Latency}

The high communication latency arises from two factors: extensive communication data volume and irregular communication patterns. To reduce latency, it can be tackled on two fronts. Firstly, data scale can be curtailed through techniques like data quantization and sampling. Concurrently, optimizing data layout via reorder can ameliorate irregular communication, ultimately enhancing the utilization of communication bandwidth.

$\bullet$ \textbf{Quantization.}
This method produces a more compact representation, reducing transmitted data amount during communication and then mitigating communication latency. This process skillfully balances data size reduction with preserving essential information for accurate computation, bolstering the efficiency of distributed graph processing. Notably, several distributed GNN frameworks have integrated quantization, including BiFeat \cite{2022BiFeat} and QGTC \cite{2022QGTC}.

$\bullet$ \textbf{Sampling.}
Through sampling mini-batches, the communication data volume between computing components can be diminished, thus reducing synchronization latency. Compared to full-batch training, sampling-based training significantly reduces communication overhead, enabling more efficient parallelism, faster convergence, and the ability to handle larger datasets. Table \ref{table:Efficient-communication} lists sampling-based parallel frameworks.

$\bullet$ \textbf{Reorder.}
This method reduces inter-node communication by grouping related vertices and edges. Reordered layouts simplify irregular communication and enhance data access efficiency, thus lowering the resources and time spent on communication and mitigating communication latency in distributed graph processing. An example is the RabbitOrder \cite{2016RabbitOrder} framework, which employs graph reorder to reduce communication overhead.

%% file: tex/comparison.tex
The preceding sections have meticulously explored GPP methods through the lenses of algorithmic principles and hardware optimization, while showcasing illustrative examples. This section provides a comprehensive summary and comparison of seven GPP methods, as outlined in Tables \ref{table:summary-algorithm} and \ref{table:summary-hardware}.

\begin{table*}[h]
\vspace{-0.3cm}  
\caption{Summary of GPP methods in algorithmic optimization}
\vspace{-0.3cm}  
\label{table:summary-algorithm}
\centering
\small
\begin{tabular}{|c|c|c|c|c|c|}
\hline
\multirow{2}{*}{\textbf{Method}} & \textbf{TGC} & \multicolumn{2}{c|}{\textbf{GNN}}       & \multirow{2}{*}{\makecell{\textbf{Information}\\ \textbf{Loss}}} & \multirow{2}{*}{\makecell{\textbf{Accuracy}\\ \textbf{Sacrifice}}}\\ 
\cline{2-4}
                        & Algorithm  & \multicolumn{1}{c|}{Backbone} & Phase            &              &    \\ 
\hline
Partition               & All       & All            & Training \& Inference              &  \checkmark  &  \ding{53}  \\
\hline
Sampling                &  -        & GCN, GAT       & Training                           &  \checkmark  &  \ding{53} \\
\hline
Sparsification          & -         & GCN, GAT, GIN  & Training \& Inference              &  \checkmark  &  \ding{53} \\
\hline
Reconstruction          & -         & GCN, GAT       & Training \& Inference              &  \ding{53}   &  \ding{53} \\
\hline
Quantization            & -         & All            & Training                           &  \checkmark  &  \checkmark \\
\hline
Reorder                 & All       & GCN, GAT       & Training \& Inference              &  \ding{53}   &  \ding{53} \\
\hline
Reorganization          & All       & GCN, GAT       & Training \& Inference              &  \ding{53}   &  \ding{53} \\
\hline
\end{tabular}
\vspace{-0.3cm}  
\end{table*}

\begin{table*}[h]
\vspace{-0.3cm}  
\caption{Summary of GPP methods in hardware optimization}
\vspace{-0.3cm}  
\label{table:summary-hardware}
\centering
\small
\begin{tabular}{|c|c|c|}
\hline
\textbf{Method}                 & \textbf{Optimization Effect}                                                                & \textbf{GFP Platform}     \\ 
\hline
\multirow{2}{*}{Parititon}      & \makecell[l]{Computing \ \ \ \ \ \ \ \ $\rightarrow$ Load Balance}                          & \multirow{2}{*}{CPU, GPU, FPGA, ASIC, PIM} \\ \cdashline{2-2}[1.5pt/1pt]
                                & \makecell[l]{Communication $\rightarrow$ Reduce Communication Frequency}                    &                            \\
\hline
\multirow{3}{*}{Sampling}       & \makecell[l]{Computing \ \ \ \ \ \ \ \ $\rightarrow$ Reduce Computation}                    & \multirow{3}{*}{CPU, GPU, FPGA, ASIC, PIM} \\ \cdashline{2-2}[1.5pt/1pt]
                                & \makecell[l]{Storage \ \ \ \ \ \ \ \ \ \ \ \ \ \ $\rightarrow$ Reduce Capacity Requirement} &   \\ \cdashline{2-2}[1.5pt/1pt]
                                & \makecell[l]{Communication $\rightarrow$ Reduce Communication Latency}                      &   \\
\hline
Sparsification                  & \makecell[l]{Computing \ \ \ \ \ \ \ \ $\rightarrow$ Reduce Computation}                    &  CPU, GPU \\
\hline
\multirow{2}{*}{Reconstruction} & \makecell[l]{Computing \ \ \ \ \ \ \ \ $\rightarrow$ Reduce Computation}                    & \multirow{2}{*}{CPU, GPU, FPGA, ASIC} \\ \cdashline{2-2}[1.5pt/1pt]
                                & \makecell[l]{Storage \ \ \ \ \ \ \ \ \ \ \ \ \ \ $\rightarrow$ Reduce Data Movement}        &   \\
\hline
\multirow{3}{*}{Quantization}   & \makecell[l]{Computing \ \ \ \ \ \ \ \ $\rightarrow$ Reduce Computation}                    &  CPU, GPU, FPGA                        \\ \cdashline{2-3}[1.5pt/1pt]
                                & \makecell[l]{Storage \ \ \ \ \ \ \ \ \ \ \ \ \ \ $\rightarrow$ Reduce Capacity Requirement} & \multirow{2}{*}{CPU, GPU}              \\ \cdashline{2-2}[1.5pt/1pt]
                                & \makecell[l]{Communication $\rightarrow$ Reduce Communication Latency}                      &   \\
\hline
\multirow{3}{*}{Reorder}        & \makecell[l]{Computing \ \ \ \ \ \ \ \ $\rightarrow$ Reduce Computation}                    & PIM  \\ \cdashline{2-3}[1.5pt/1pt]
                                & \makecell[l]{Storage \ \ \ \ \ \ \ \ \ \ \ \ \ \ $\rightarrow$ Reduce Data Movement}        & \multirow{2}{*}{CPU, GPU, FPGA, ASIC}  \\ \cdashline{2-2}[1.5pt/1pt]
                                & \makecell[l]{Communication $\rightarrow$ Reduce Communication Latency}                      &   \\
\hline
Reorganization                  & \makecell[l]{Storage \ \ \ \ \ \ \ \ \ \ \ \ \ \ $\rightarrow$ Reduce Data Movement}        & CPU, GPU, FPGA, ASIC, PIM  \\
\hline
\end{tabular}
\vspace{-0.4cm}  
\end{table*}

$\bullet$ \textbf{TGC vs. GNN:} GPP methods for TGC leverage graph characteristics to optimize performance. Among these methods, partition evenly divides large-scale graphs, while reorder and reorganization enhance data access, considering factors like graph irregularity and the power-law distribution. Importantly, partition, reorder, and reorganization are versatile techniques not bound to specific TGC algorithms. These well-established GPP methods have the potential to optimize the execution of all TGC algorithms.

GPP methods for GNNs take into account the specific characteristics of GNN models. Among these methods, reconstruction aims to eliminate redundancy in aggregation, thereby enhancing data utilization. Other methods are derived from technologies used in TGC, such as partition, reorder, and reorganization, or from NNs, including sampling, sparsification, and quantization. This integration is due to GNNs combining graph structures and NNs. These methods are adapted and optimized to align with the unique requirements of GNNs. Notably, sampling and quantization are GPP methods designed explicitly to accelerate GNN training, while other GPP methods can serve both training and inference phase.
Nobably, a significant optimization efforts are directed towards ConvGNN models like GCN, GAT, and GIN. Consequently, technologies such as sampling, sparsification, reconstruction, reorder, and reorganization are closely integrated with convolutional models, specifically tailored to meet their unique requirements. While partition and quantization are general GPP methods, their primary application remains the optimization of GCN and GAT execution.

$\bullet$ \textbf{Information Loss vs. Accuracy Sacrifice:} Reconstruction, reorder, and reorganization do not incur information loss because they solely change the storage access order of data. In contrast, other GPP methods can result in information loss. These methods include partition, sampling, and sparsification, which involve the removal of edges and vertices, and quantization, which involves precision degradation. It's important to highlight that only quantization techniques are typically associated with potential accuracy sacrifice, since it significantly compresses the precision of data used for training. Fortunately, given the inherent robustness of NNs, this accuracy sacrifice can often be managed within an acceptable range through refinement and adjustment. 

Conversely, while partition may cut vertices and edges, the resulting information loss is generally minimal, preserving accuracy. Sampling methods selecting subgraphs for mini-batches inherently incur information loss, yet well-designed strategies can retain vital structural and feature-related details while still maintaining accuracy. Sparsification, which selectively removing edges from the original graph, inherently results in some loss of information,but it can serve as a data augmentation strategy to alleviate overfitting in deep GNNs and thereby improving training accuracy.

$\bullet$ \textbf{Optimization Effect vs. GFP Platform:} Partition and sampling are versatile GPP methods that accelerate a range of graph processing algorithms and yield varied hardware optimization benefits. Among these, partition divides large graphs into manageable subgraphs for parallel processing, aiming for load balance and reduced communication frequency. Balancing these goals presents challenges, requiring trade-offs based on the application. On the other hand, sampling delivers optimization in computing, storage, and communication, particularly for large-scale GNN training. It reduces per-iteration computation, accelerates convergence through smaller subgraphs, and trims memory and communication overhead.  Sampling-based mini-batch training is favored for its efficiency and scalability in large-scale GNN scenarios. Notably, many large-scale GNN training scenarios adopt the partition+sampling approach. Careful consideration of sampling locality and overhead is essential during partitioning, which influences method selection.

Sparsification reduces computational demands, speeding up execution by removing redundant edges. However, in GNN training, many sparsification techniques prioritize accuracy improvements. While reducing per-iteration time through data reduction, the elevated iteration count for convergence might lead to an overall runtime increase. Consequently, existing GNN works are mostly using sparsification for accuracy enhancement and are commonly deployed on general platforms such as CPUs and GPUs.

Graph reconstruction proves efficient in curtailing redundant computations and data movements through the reuse of intermediate results. This approach involves minimizing repetitive operations and optimizing the utilization of computing resources, which holds particular advantages for customized architectures, resulting in substantial resource conservation. In heterogeneous architectures, where pre-processing occurs on CPUs, graph reconstruction helps mitigate data transfers between CPUs and specialized platforms like FPGAs and ASICs, further hastening GNN execution. On general platforms such as CPUs and GPUs, graph processing frameworks store reusable vertices in cache and directly access previously calculated results when recalculation is required, thereby achieving acceleration.

Quantization offers a notable advantage through significant memory reduction achieved via data compression, a feature particularly beneficial for memory-constrained edge devices. Moreover, data compression contributes to the reduction of communication overhead in distributed GNN computation, since transmitting quantized data across nodes requires fewer bits, leading to decreased communication latency and alleviating network congestion. As a result, quantization is primarily employed in CPU or GPU-based large-scale graph processing frameworks. Furthermore, representing data with fewer bits simplifies complex floating-point operations into shifts, conserving computing resources and energy, especially when applied to FPGAs.

Reorder enhances data locality by grouping frequently accessed data together, minimizing the need for retrieval from distant memory locations and unnecessary data movement. This optimization is particularly beneficial when irregular memory access patterns lead to high cache misses and inefficient data transfer. In parallel graph processing, reorder improves data locality, reduces irregular node communication, and decreases communication latency. Reorder is applicable across various platforms, including CPU, GPU, FPGA, and ASIC. Additionally, for PIM-based architectures, reorder enhances computing efficiency for ReRAM by reducing zero elements in matrices.

Reorganization is a fundamental data flow optimization technique employed to enhance memory access efficiency. It achieves this by transforming irregular memory accesses into contiguous ones, effectively reducing data movement resulting from non-sequential data access. This optimization technique finds widespread use in graph processing implementations across a range of platforms, including CPU, GPU, FPGA, ASIC, and PIM.

%% file: tex/future.tex
In previous sections, we extensively discussed various GPP methods from algorithmic and hardware standpoints, demonstrating their efficacy in accelerating graph processing applications. However, challenges persist due to algorithmic and device-related factors. In this section, we explore challenges within GPP and discuss potential future directions.

\subsection{Challenges of GPP}

Currently, the field of GPP faces four primary challenges:

$\bullet$ \textbf{High Overhead}: As data scales continue to increase, the time overhead required for GPP becomes more pronounced. For instance, graph reorder time that takes 13.4 seconds on the Pokec dataset can take as long as 1.5 hours on the Twitter dataset \cite{2016Gorder}. While certain heuristics like Hub-Sorting and Hub-Clustering have been introduced to reduce reorder time, these methods may fall short compared to more intricate techniques, potentially affecting result accuracy. Therefore, there is an urgent need to develop efficient and optimized GPP approaches tailored for large-scale graphs.

$\bullet$ \textbf{Dynamic Graphs}: Dynamic graph processing has gained popularity \cite{i2020dynamicGCN, 2022dynamicGNN-recommdation}, and GPP techniques designed for static graphs may not directly apply to dynamic graphs that evolve over time. More online GPP techniques that can adapt to dynamic graphs and minimize the GPP overhead are needed. In scenarios where dynamic graph data arrives in a streaming fashion, GPP methods must efficiently process and update the graph representation on-the-fly. This requires designing algorithms that can handle the data stream efficiently and make real-time decisions.

$\bullet$ \textbf{Accuracy Loss}: GPP methods such as quantization and sampling are commonly employed to improve execution efficiency, but they can introduce precision loss \cite{2021EXACT, 2021liu-sampling-survey}. Although advancements have been made in mitigating accuracy loss caused by these techniques, achieving a balance between accuracy and optimization remains a challenge. Future research endeavors should strive to discover innovative GPP methods capable of minimizing precision loss while achieving substantial computational gains. This will enable more effective improvement of overall performance.

$\bullet$ \textbf{Heterogeneous Graphs}: While existing GNN methods predominantly focus on isomorphic graphs, the recent emergence of heterogeneous graph models has sparked a new research trend \cite{2020Hgcn, 2022sybilflyover-hetero, 2022heteroGNNgate, 2022survey-hetero}. As heterogeneous graphs consist of different types of edges and vertices \cite{2018meta-path}, GPP techniques designed for isomorphic graphs are not directly applicable. To effectively analyze and process the complex graph structures represented by heterogeneous graphs, it is essential to develop tailored GPP techniques specifically designed for such graphs.

\subsection{Future Prospects of GPP}


\hspace{1em} $\bullet$ \textbf{GPP Profiling}: Addressing the aforementioned challenges requires a thorough assessment of the behavioral characteristics and performance bottlenecks of graph processing and pre-processing. This evaluation provides crucial insights for optimizing execution. Previous efforts have evaluated some aspects, such as GCN execution and distributed GNN training on GPUs \cite{2020Characterizing-GCN-GPU, 2022characterizing-distributed-GNNtraining-GPU}, as well as bottlenecks in dynamic and heterogeneous graphs \cite{2022bottleneckDGNN, 2022characterizing-HGNN-GPU}. Nonetheless, a comprehensive analysis and evaluation of GPP remain unexplored, presenting a promising avenue for future research.

$\bullet$ \textbf{Pipeline-friendly GPP}: With the expansion of data scales and the escalation of model complexity, the impact of GPP overhead becomes more pronounced, particularly in the realm of online methods. While more intricate GPP algorithms might offer better acceleration, it could potentially introduce pipeline stalls within the entire system. Thus, achieving a delicate trade-off between the provided acceleration and GPP overhead is of paramount importance. An effective approach involves the discerning selection of appropriate GPP methods and the meticulous optimization of the temporal overlap between GPP and GFP steps to attain optimal acceleration. For instance, GraphACT \cite{2020GraphACT} integrates a swift graph reconstruction technique for online GPP, seamlessly integrating with GNN inference execution on FPGA. Furthermore, efforts are channeled towards the improvement of GPP algorithmic execution; for example, GNNSampler \cite{2021gnnsampler} implements locality-aware optimizations to sampling algorithms, aimed at curtailing sampling overhead.

$\bullet$ \textbf{Accuracy-friendly GPP}: Certain GPP methods have the potential to impact the final accuracy of graph processing. Some works make efforts to mitigate this issue and reduce the influence of GPP on accuracy. For instance, works like SGQuant \cite{2020SGQuant} and DegreeQuant \cite{2020DegreeQuant} have developed degree-based mechanisms for quantization, specifically designed to minimize precision loss. Another approach, known as EXACT \cite{2021EXACT}, addresses precision loss through projection techniques. It is important to consider that some of these methods may require additional memory space. Striking a balance between accuracy loss and optimization effects while carefully considering the trade-offs involved is crucial in achieving both accurate and efficient graph processing.

$\bullet$ \textbf{Comprehensive Framework of GPP}: Optimizing GPP often involves combining multiple GPP methods. In distributed GNN training systems, as discussed in \cite{2020DistDGL} and \cite{2020PCGCN}, input graphs are partitioned into subgraphs and sampled for training. Others, such as \cite{2020GcnInference} and SnF \cite{2022SnF}, utilize reconstruction, reorder, and sampling. Exploring the adaptability and potential conflicts between various GPP methods is crucial for developing a comprehensive and configurable GPP framework. A recent example is EndGraph \cite{2022EndGraph}, a distributed pre-processing framework that accelerates graph partition and construction, improving pre-processing performance up to 35.76× compared to state-of-the-art systems.

$\bullet$ \textbf{Specific Accelerators for GPP}: GPP is typically performed on CPUs, even when executing formal processing on GPUs or custom architectures. For instance, CuSha \cite{2014cusha} performs partition and reorganization on CPUs before executing traditional graph computing on GPUs. Similarly, heterogeneous platforms like GraFBoost \cite{2018grafboost} and GraphACT \cite{2020GraphACT} utilize CPU-based GPP tasks before executing graph algorithms on FPGA. However, due to the increasing GPP overhead, it is worthwhile to design dedicated graph GPP accelerators to enhance online GPP and facilitate seamless collaboration with fast-executing rear accelerators. For example, I-GCN \cite{2021I-gcn} has developed specialized architectures for graph reconstruction, resulting in significant improvements in overall processing speed. Implementing such accelerators, GPP tasks can be efficiently handled in real-time, effectively improving the system performance.

$\bullet$ \textbf{GPP for Heterogeneous Graphs}: With the increasing popularity of heterogeneous graphs, the optimization of their processing has emerged as a significant research area. Despite existing efforts towards optimizing the execution of such graphs, exemplified by accelerators like HiHGNN \cite{2023HiHGNN}, there remains a demand for specialized GPP techniques catering to these complex structures. While a few GPP methods designed for heterogeneous graphs already exist, such as HetGNN \cite{2019HetGNN} and HGSampling \cite{2020HGSampling}, further research is essential to develop more advanced GPP approaches.

%% file: tex/conclusion.tex

Graph pre-processing (GPP) is a crucial step that transforms raw graph data to prepare it for the formal execution of graph processing algorithms. GPP is widely used across various execution systems, including general frameworks on CPUs and GPUs, as well as custom accelerators based on FPGA, ASIC, PIM, and more. However, as graph data scales up, the overhead of GPP becomes significant, necessitating thorough analysis and optimization of GPP.

In this paper, we present a comprehensive survey of GPP methods from both algorithmic and hardware perspectives, aiming to provide valuable insights for optimizing graph processing algorithms and hardware acceleration. We discuss the challenges in graph processing execution and emphasize the critical role of GPP in graph processing. Existing GPP methods can be concluded into seven types: partition, sampling, sparsification, reconstruction, quantization, reorder, and reorganization. Our taxonomy of GPP methods comprises two decision levels, where algorithmic categories include graph representation and data representation optimization, while hardware categories encompass efficient computation, storage, and communication.

Finally, we provide a summary and comparison of GPP methods, discussing challenges and potential future directions. Despite facing challenges such as high overhead and adaptability concerns, certain endeavors have been undertaken to enhance GPP. Many GPP techniques hold an irreplaceable position in heightening the efficiency of various graph applications on diverse platforms. We await increased attention and exploration of GPP.

%% file: tex/acknowledgement.tex
This work was supported by the National Key Research and Development Program under Grant 2022YFB4501404, in part by the National Natural Science Foundation of China under Grant 62202451, the CAS Project for Young Scientists in Basic Research under Grant YSBR-029, and the CAS Project for Youth Innovation Promotion Association. 